\documentclass[reprint,showpacs,preprintnumbers,amsmath,amssymb,prc,floatfix]{revtex4-1}
\usepackage{float}
\usepackage{color}
\usepackage{graphicx}
\usepackage{dcolumn}
\usepackage{bm}
\usepackage{xcolor}
\usepackage{comment}
\usepackage[colorlinks,citecolor=blue,linkcolor=red,anchorcolor=blue,filecolor=blue,urlcolor=blue]{hyperref}
\usepackage{multirow}
\usepackage{threeparttable} 
\usepackage{adjustbox} 
\usepackage{color}
\usepackage{longtable}

\begin{document}


\title{Systematic study of fusion barrier characteristics within the relativistic mean-field formalism}

\author{Shilpa Rana$^1$}
\email{srana60\_phd19@thapar.edu}
\author{M. Bhuyan$^{2}$}%
\email{bunuphy@um.edu.my}
\author{Raj Kumar$^1$}
\email{rajkumar@thapar.edu}
\bigskip
\affiliation{$^1$School of Physics and Materials Science, Thapar Institute of Engineering and Technology, Patiala 147004, India}
\affiliation{$^2$Department of Physics, Faculty of Science, University of Malaya, Kuala Lumpur 50603, Malaysia}

\date{\today}
\bigskip 
\begin{abstract}
\noindent
{\bf Background:} Heavy-ion fusion reactions play a crucial role in various aspects of nuclear physics and astrophysics. The nuclear interaction potential and hence the fusion barrier formed between the interacting nuclei are the keys to understanding the complex fusion process dynamics. Thus, a theoretical investigation of fusion barrier characteristics which includes the relativistic effects is of paramount significance. \\   
\noindent
{\bf Purpose:} This work intends to explore the fusion barrier characteristics of different target-projectile combinations within the relativistic mean-field (RMF) formalism. The M3Y nucleon-nucleon (NN) potential is compared with the relativistic R3Y and density dependent R3Y (DDR3Y) NN potentials within the double folding approach. A systematic study is carried out to study the effect of different RMF density distributions and effective NN interactions on the fusion and/or capture cross-section of twenty-four (24 nos) target-projectile combinations leading to heavy and superheavy nuclei (SHN).

\noindent 
{\bf Methods:} The density distributions of interacting nuclei and the microscopic R3Y NN interaction are obtained from relativistic mean-field (RMF) formalism for non-linear NL1, NL3, TM1, and relativistic-Hartree-Bogoliubov (RHB) approach for DDME2 parameter sets. The medium independent relativistic R3Y, the density dependent DDR3Y, and widely adopted M3Y NN potential are used to obtain the nuclear interaction potential within the double folding approach. The densities for the projectiles and targets are obtained from the relativistic mean-field approaches. The fusion and/or capture cross-section for the different reaction systems is calculated using the well-known $\ell$-summed Wong model. \\
\noindent
{\bf Results:} The barrier height and position of 24 heavy-ion reaction systems are obtained for different nuclear density distributions and effective NN interaction potentials. We have considered the lighter mass projectile and heavier mass target combinations for synthesizing exotic drip-line nuclei, including the super-heavy nuclei (SHN). These reactions include the even-even $^{48}$Ca+$^{154}$Sm, $^{48}$Ca+$^{238}$U, $^{48}$Ca+$^{248}$Cm, $^{26}$Mg+$^{248}$Cm; even-odd $^{46}$K+$^{181}$Ta; odd-odd $^{31}$Al+$^{197}$Au and $^{39}$K+$^{181}$Ta and also seventeen other systems for the synthesis of SHN Z=120. The comparison of fusion and/or capture cross-section obtained from the $\ell$-summed Wong model is made with the available experimental data. \\
\noindent
{\bf Conclusions:} The phenomenological M3Y NN potential is observed to give higher barrier heights than the relativistic R3Y NN potential for all the reaction systems. The comparison of results obtained from different relativistic parameter sets shows that the densities from NL1 and TM1 parameter sets give the lowest and highest barrier heights for all the systems under study. The density dependant DDR3Y NN potential is obtained within the Relativistic-Hartree-Bogoliubov approach for the DDME2 parameter set. We observed higher barrier heights and lower cross-sections for DDR3Y NN potential as compared to density-independent R3Y NN potentials obtained for considered non-linear NL1, NL3 and TM1 parameter sets. According to the present analysis, it is concluded that the NL1 and NL3 parameter sets provide comparatively better overlap with the experimental fusion and/or capture cross-section than the TM1 and DDME2 parameter sets.
\end{abstract}
\maketitle

\section{INTRODUCTION}
\noindent
The study of underlying physics involved in low energy heavy-ion fusion reactions is essential for a better understanding of the characteristics of nuclear forces, nuclear structure, superheavy nuclei (SHN), magic shell closure, drip lines, and other related phenomena  \cite{mont17,cheng19,canto20,toub17,wakhle18,das98,litn20,kozu16,ghor20,sahoo20}. The interaction potential and consequently the fusion barrier formed between the projectile and target nuclei provides the basis to understand the dynamics involved in these fusion reactions. The characteristics of the fusion barrier such as the barrier height, position and oscillator frequency etc. are used further to calculate one or other related physical quantities such as the fusion probability and cross-section \cite{manj18,dutt10,ghodsi13}. Since the fusion barrier is not the direct measurable quantity in the experiments, so theoretical modelling is required to extract its characteristics \cite{cant06}. The origin of the fusion barrier is the interplay between the attractive short-range nuclear potential and the repulsive long-range Coulomb potential. The Coulomb potential is a known quantity and has a well-established formula. For the calculation of nuclear potential, we have different theoretical approaches available in literature \cite{raj11,deep13,rajk11,deep14,back14,bloc77,bloc81,cheng19,bass73,bass74,bass77,vaut72,raj20}. These theoretical models differ from each other in respect of their basic assumptions and the parameters used. As a consequence, the fusion barrier characteristics calculated also vary and greatly depend upon the adjustments of parameters used in a theoretical formalism \cite{raj11,deep13,rajk11,deep14,back14,bloc77,bloc81,cheng19,bass73,bass74,bass77,vaut72,raj20}. \\ \\
The phenomenological proximity potentials based on the proximity theorem \cite{bloc77,bloc81} are widely used to estimate the nuclear interaction potential in terms of the mean curvature of the interacting surfaces and a universal function of separation distance  \cite{bloc77,bloc81,raj11,deep13,cheng19}. The Bass potential based upon the liquid drop model also provides a simple exponential form for the nuclear interaction potential \cite{cheng19,bass73,bass74,bass77}. Moreover, the semi-microscopic approaches describe nuclear interaction potential as the difference in the energies of interacting nuclei at an infinite separation and a distance when overlapping. The examples are asymmetric two-center shell model and the models based on the energy density formalism (EDF) \cite{rajk11,deep14,dutra12,maru72,gram79,gram80,brac85,vaut72}. Furthermore, the phenomenological double folding approach also has successful applications to deduce the interaction potential between the two colliding heavy-ions \cite{bert79,satc79,khoa16,long08}. In this approach, the nuclear optical potential is obtained using nuclear density distributions and effective nucleon-nucleon interaction. The model has been widely adopted to provide real and imaginary parts of the optical potentials between the colliding ions in elastic and inelastic scattering as well as in the study of nuclear fusion characteristics \cite{bert79,satc79,bhuy18,bhuy20,khoa16,long08} and references therein. \\ \\
Nuclear fusion is considered to be a complex phenomenon since a large number of nucleons are involved. A complete description of nucleon-nucleon (NN) interaction potential is a prerequisite to understanding these nucleon's interactions. One of the well-known approaches to handle the nuclear many-body problem is the self-consistent relativistic mean-field model. Recently, the microscopic R3Y NN interaction potential has also been derived from the relativistic mean-field formalism \cite{sing12,sahu14,lahi16}. In relativistic mean-field (RMF) models, the point-like Dirac nucleons interact through the exchange of mesons and photons \cite{ring96,vret05,meng16,bodmer91,rein86,lala97,bhuy18,bhuy12,bhuy20}. The mass of $\sigma$ meson and the coupling constants of the interacting mesons are fine-tuned to the ground state bulk properties of finite nuclei. A number of these parameter sets with linear and/or non-linear meson couplings (e.g. HS, models with name starting with L and NL, TM1, FSUGold, IOPB-I, G2 and G3 etc.),  density-dependent meson-exchange couplings (DDME1, DDME2, DDME$\delta$ etc.) and zero-range point-couplings (DD-PC1, PC-PK1 etc.) \cite{rein86,lala97,bodmer91,sing12,ring96,vret05,meng16,bhuy20,tani20,gupta97,kumar17,kumar18,niks02,type99,fuchs95,furn97} and references therein are available. All these defined parameter sets provide overall satisfactory results for the nuclear bulk properties and also explain the nuclear matter observables \cite{rein86,lala97, meng06,bhuy12,zhang05,afan96,afan96A,bodmer91,kumar17,kumar18,niks02,lala05,type99,fuchs95,hofm01,bhuy15,bhu18,kumar19}. With the frequent measurements of various bulk properties from experiments and the constraints on nuclear matter observables including highly dense and isospin asymmetric systems, all these relativistic parameters are developed. Each parametrization in the relativistic mean-field model has its own identities and certain limitations, for more details follow Ref. \cite{dutra14}. In the present study, we have considered two different kinds of parametrizations, such as non-linear NL1, NL3, and TM1 parameter sets within the relativistic mean-field model and density-dependent DDME2 parameter set within Relativistic-Hartree-Bogoliubov (RHB) approach to study the characteristics of nuclear fusion in terms of the nuclear density distributions and NN potential. The medium dependent R3Y (named as DDR3Y) NN potential given in terms of density-dependent nucleon-meson couplings was introduced for the first time in the fusion study and can be found in Ref. \cite{rana22}. Along with the relativistic R3Y and DDR3Y potentials, we have also employed the widely adopted M3Y potential to estimate the nucleon-nucleon interaction potential \cite{satc79,bert79} for the comparison.\\ \\
A systematic analysis will be carried out in three steps to study the effect of nucleon-nucleon interaction potential and the density distributions of the fusing nuclei on the fusion barrier characteristics and consequently on the fusion and/or capture cross-section. In the first step, a comparison will be made for the widely used M3Y and recently developed relativistic R3Y effective nucleon-nucleon (NN) potential in terms of nuclear potential within the double folding approach. Moreover, the density-dependent R3Y (DDR3Y) NN interaction potential obtained for the DDME2 parameter set within the RHB approach is also taken into account in this analysis. In the second step, the effect of RMF nuclear density distributions obtained for non-linear NL1, NL3, TM1, and density-dependent DDME2 parameter sets will be analyzed on the fusion characteristics. Finally, in the third step, we will study the effect of R3Y NN potential obtained for non-linear NL1, NL3, and TM1 parameter sets within the RMF model as well as the DDR3Y NN potential obtained for the density-dependent DDME2 parameter set within the RHB approach on the fusion barrier characteristics and consequently on the fusion and/or capture cross-section. We have chosen 24 different light mass projectile and heavy mass target combinations from the various exotic regions of the nuclear chart in the present analysis. The even-even $^{48}$Ca+$^{154}$Sm, $^{48}$Ca+$^{238}$U, $^{48}$Ca+$^{248}$Cm, $^{26}$Mg+$^{248}$Cm; even-odd $^{46}$K+$^{181}$Ta; and odd-odd $^{31}$Al+$^{197}$Au and $^{39}$K+$^{181}$Ta systems of lighter mass projectile and heavier mass target nuclei. These reaction systems involve neutron-rich projectiles and are pertinent for the synthesis of neutron-rich heavy, and superheavy nuclei \cite{wata01,knya07,itki04,prok05,kozu10,wakhle18}. Seventeen different possible systems, namely, $^{40}$Ca + $^{257}$Fm, $^{48}$Ca + $^{254}$Fm, $^{46}$Ti + $^{248}$Cf, $^{46}$Ti + $^{249}$Cf, $^{50}$Ti + $^{249}$Cf, $^{50}$Ti + $^{252}$Cf, $^{50}$Cr + $^{242}$Cm, $^{54}$Cr + $^{248}$Cm, $^{58}$Fe + $^{244}$Pu, $^{64}$Ni+$^{238}$U, $^{64}$Ni + $^{235}$U, $^{66}$Ni + $^{236}$U, $^{50}$Ti + $^{254}$Cf, $^{54}$Cr + $^{250}$Cm, $^{60}$Fe + $^{244}$Pu, $^{72}$Zn + $^{232}$Th and $^{76}$Ge + $^{228}$Ra are chosen leading to the synthesis of different isotopes of highly discussed superheavy Z = 120 \cite{bhuy12,adam09,hoff04}. It is worth mentioning that all these isotopes are neutron-rich and also $^{304}120$ is predicted to be a doubly magic shell nuclei within various theoretical models \cite{bhuy12,zhang05,agbe15,lala96,rutz97,sobi66,nils69,adam09}. Hence, it would be of interest to study the effect of different nuclear density distributions and nucleon-nucleon interaction potentials on the fusion and/or capture cross-section for these systems. The $\ell$-summed Wong model \cite{kuma09} is employed to deduce the fusion and/or capture cross-section, and comparison with the experimental cross-section is also made wherever available. \\ \\
The paper is organized as follows: the theoretical formalism for nuclear potential using the relativistic mean-field approach and the double folding procedure is explained in section \ref{theory} along with a brief description of the $\ell-$summed Wong model to estimate the fusion and/or capture cross-section. The results obtained from the calculations are discussed in section \ref{results}. In Section \ref{summary} summary and conclusions of the present work are made. \\

\section{Nuclear Interaction Potential from Relativistic Mean Field Formalism}
\label{theory}
A complete description of the total interaction potential is crucial to estimate the fusion probability of two colliding nuclei. This interaction potential comprises three parts: the nuclear interaction potential, the Coulomb potential, and the centrifugal potential. The total interaction potential ($V_T^{\ell}(R)$) between the projectile and target nuclei can be written as,
\begin{eqnarray}
V_{T}^{\ell}(R)= V_n (R)+V_C(R)+ V_{\ell}(R).
\label{vtot}
\end{eqnarray}
Here, $V_C(R)=Z_pZ_te^2/R$ and $V_{\ell}(R)=\frac{\hbar^2\ell(\ell+1)}{2\mu R^2}$ are the Coulomb and centrifugal potential, respectively. $\mu$ is the reduced mass, and $R$ is the separation distance. The nuclear potential $V_n (R)$ is calculated here within the double folding approach \cite{satc79},
\begin{eqnarray}
V_{n}(\vec{R}) & = &\int\rho_{p}(\vec{r}_p)\rho_{t}(\vec{r}_t)V_{eff}
\left( |\vec{r}_p-\vec{r}_t +\vec{R}| {\equiv}r \right) \nonumber \\
&& d^{3}r_pd^{3}r_t .
\label{fold}
\end{eqnarray}
Here,  $\rho_p$ and $\rho_t$ are the total density (sum of proton and neutron densities) distributions of projectile and target nuclei, respectively. $V_{eff}$ is the effective nucleon-nucleon (NN) interaction. There are several expressions for the effective NN interaction potential available in the literature. One of the well-known expressions is known as the M3Y (Michigan 3 Yukawa) potential \cite{bert79}. As the name suggests, it consists of three Yukawa terms \cite{bert79,satc79,bhuy18,lahi16} and is given by,
 \begin{eqnarray}
V_{eff}^{M3Y}(r)= 7999 \frac{e^{-4r}}{4r}-2140\frac{e^{-2.5r}}{2.5r}+J_{00}(E)\delta(r). 
\label{m3y}
\end{eqnarray}
Here, $J_{00}(E)\delta(r)$ is long-range one pion exchange potential because there can be the possibility of nucleon exchange between the projectile and target nuclei.
 
As mentioned before, the characteristics of the total interaction potential play a crucial role in determining one or another fusion properties. The barrier characteristics such as the barrier height ($V_B^\ell$) and barrier position ($R_B^\ell$)  can be determined from Eq. (\ref{vtot}) using following conditions:
 \begin{eqnarray}
 \frac {dV_{T}^{\ell}}{dR} \bigg |_{R=R_{B}^{\ell}}=0.
 \label{vb1}
 \end{eqnarray}
 \begin{eqnarray}
 \frac{d^2V_{T}^{\ell}}{dR^2}\bigg |_{R=R_{B}^{\ell}}\le 0.
 \label{vb2}
 \end{eqnarray}
 Moreover, the barrier curvature ($\hbar\omega_\ell$) is evaluated at $R = R_{B}^{\ell}$ corresponding to the barrier height $V_{B}^{\ell}$, and given as,
 \begin{eqnarray}
 \hbar \omega_{\ell}=\hbar [|d^2V_{T}^{\ell}(R)/dR^2|_{R=R_{B}^{\ell}}/\mu]^{\frac{1}{2}}.
 \label{vb3}
\end{eqnarray}
To determine these quantities, we need a complete description of nuclear interaction potential, which is calculated here using the double folding integral given by Eq. (\ref{fold}). The main requirements to solve the double folding integral are the nuclear density distributions and effective nucleon-nucleon interaction potential. The well-known relativistic mean-field (RMF) formalism and relativistic-Hartree-Bogoliubov (RHB) have been employed to determine these density distributions and NN interaction potential. The RMF formalism has its successful applications in describing the properties of nuclear matter as well the finite nuclei \cite{ring96,bodmer91,rein86,lala97,bhuy18,bhuy12,bhuy20,meng06,tani20,vret05,gupta97} . A phenomenological description of nucleon interaction through the exchange of mesons and photons is given by RMF Lagrangian density \cite{ring96,bodmer91,rein86,lala97,bhuy18,bhuy12,bhuy20,meng06,tani20,vret05,gupta97} which can be written as,
\begin{eqnarray}
{\cal L}&=&\overline{\psi}\{i\gamma^{\mu}\partial_{\mu}-M\}\psi +{\frac12}\partial^{\mu}\sigma
\partial_{\mu}\sigma \nonumber -{\frac12}m_{\sigma}^{2}\sigma^{2}\\
&& -{\frac13}g_{2}\sigma^{3} -{\frac14}g_{3}\sigma^{4}
-g_{\sigma}\overline{\psi}\psi\sigma \nonumber -{\frac14}\Omega^{\mu\nu}\Omega_{\mu\nu}\\
&& +{\frac12}m_{w}^{2}\omega^{\mu}\omega_{\mu}+{\frac14}\xi_3(\omega^{\mu}\omega_{\mu})^2
-g_{w}\overline\psi\gamma^{\mu}\psi\omega_{\mu} \nonumber \\
&&-{\frac14}\vec{B}^{\mu\nu}.\vec{B}_{\mu\nu}+\frac{1}{2}m_{\rho}^2
\vec{\rho}^{\mu}.\vec{\rho}_{\mu} -g_{\rho}\overline{\psi}\gamma^{\mu}
\vec{\tau}\psi\cdot\vec{\rho}^{\mu}\nonumber \\
&&-{\frac14}F^{\mu\nu}F_{\mu\nu}-e\overline{\psi} \gamma^{\mu}
\frac{\left(1-\tau_{3}\right)}{2}\psi A_{\mu}.
\label{lag}
\end{eqnarray}
Here $\psi$ denotes the Dirac spinor for the nucleons, i.e., proton and neutron. $m_\sigma$, $m_\omega$ and $m_\rho$ signify the masses of isoscalar scalar $\sigma$, isoscalar vector $\omega$ and isovector vector $\rho$ mesons, respectively which intermediate the interaction between the nucleons having mass M. $g_\sigma$, $g_\omega$ and $g_\rho$ are the linear coupling constants of the respective mesons whereas the $g_2$, $g_3$ and $\xi_3$ are the non-linear self interaction constants for scalar $\sigma$ and vector $\omega$-mesons, respectively. These mass of $\sigma$ meson and the coupling constants of mesons are fitted to match the infinite nuclear matter's saturation properties and the bulk properties of magic shell nuclei.  For the present study, we have considered three parameter sets, namely NL1 \cite{rein86}, NL3 \cite{lala97}, and TM1 \cite{bodmer91}. In NL1 and NL3 parameters, only the $\sigma-$meson self-coupling non-linear terms (i.e, associated coupling constants, $g_2$ and $g_3$) are taken into account. In the case of the TM1 parameter set, the self-coupling term of vector $\omega-$meson ($\xi_3$) is considered. The terms $\tau$ and $\tau_3$ in Eq. (\ref{lag}) symbolize the isospin and and its third component, respectively. $\Omega^{\mu\nu}$, $\vec B^{\mu\nu}$ and $F^{\mu\nu}$ are the field tensors for $\omega$, $\rho$ and photons, respectively and are given as,
\begin{eqnarray}
\Omega_{\mu\nu} = \partial_{\mu} \omega_{\nu} - \partial_{\nu} \omega_{\mu},\\
\vec{B}^{\mu \nu}=\partial_{\mu} \vec{\rho}_{\nu} -\partial_{\nu} \vec{\rho}_{\mu},\\
F^{\mu\nu} =\partial_{\mu} A_{\nu}-\partial_{\nu} A_{\mu}.
\end{eqnarray}
The quantity $A_\mu$ here denotes the electromagnetic field, and the arrows symbolize the vectors in the isospin space. The equations of motion for the Dirac nucleon and the mesons are obtained from the Lagrangian density given in Eq. (\ref{lag}) using the Euler-Lagrange equations under mean-field approximation. The field equations for nucleons and mesons ($\sigma$, $\omega$, $\rho$) mesons are given as, 
\begin{eqnarray}
&& \Bigl(-i\alpha.\bigtriangledown+\beta(M+g_{\sigma}\sigma)+g_{\omega}\omega+g_{\rho}{\tau}_3{\rho}_3 \Bigr){\psi} = {\epsilon}{\psi}, \nonumber \\
&& \left( -\bigtriangledown^{2}+m_{\sigma}^{2}\right) \sigma(r)=-g_{\sigma}{\rho}_s(r)-g_2\sigma^2 (r) - g_3 \sigma^3 (r),\nonumber  \\ 
&& \left( -\bigtriangledown^{2}+m_{\omega}^{2}\right) \omega(r)=g_{\omega}{\rho}(r)-\xi_3\omega^3(r),\nonumber   \\  
&& \left( -\bigtriangledown^{2}+m_{\rho}^{2}\right) \rho(r)=g_{\rho}{\rho}_3(r).
\label{field}
\end{eqnarray} 
It is to be noted here that the terms with $\sigma^3$ and $\sigma^4$  account for the self-coupling among the scalar sigma mesons. Similarly, the term with $\omega^4$ takes care of self-coupling among the vector mesons. These non-linear self-coupling terms take care of saturation properties and also soften the equation of state of the nuclear matter \cite{sahu14,rein86, bodmer91,lala97}. The properties of finite nuclei such as the binding energy and charge radius ($r_{ch}$) estimated from the Lagrangian density containing the non-linear $\sigma-\omega$ terms also give satisfactory match with the experimental values \cite{ring96,bodmer91,rein86,lala97,meng06,vret05,gupta97}. The alternative approach to introduce the density-dependent nucleon-meson couplings within relativistic mean field is the relativistic-Hartree-Bogoliubov (RHB) approach \cite{niks02,lala05,type99,fuchs95,hofm01}. In this framework, the couplings of $\sigma$, $\omega$ and $\rho$ mesons to the nucleon fields (i.e. $g_\sigma$, $g_\omega$ and $g_\rho$) are defined as \cite{niks02,lala05,type99,fuchs95,hofm01},
\begin{eqnarray}
g_i(\rho)=g_i(\rho_{sat})f_i(x)|_{i=\sigma,\omega},
\label{dd1}
\end{eqnarray}
where
\begin{eqnarray}
f_i(x)=a_i\frac{1+b_i(x+d_i)^2}{1+c_i(x+d_i)^2}
\label{dd2}
\end{eqnarray}
and
\begin{eqnarray}
g_\rho(\rho)=g_\rho(\rho_{sat})exp[-a_\rho(x-1)].
\label{dd3}
\end{eqnarray}
Here, $x=\rho / \rho_{sat}$, with $\rho_{sat}$ is the baryon density of symmetric nuclear matter at saturation. The five constraints- $f_i(1)=1$, $f''_i(0)=0$, and $f''_\sigma(1)=f''_\omega(1)$ reduce the number of independent parameters in Eq. (\ref{dd2}) from eight to three. All the independent parameters (the mass of $\sigma$ meson and coupling parameters) are obtained to fit the ground state properties of finite nuclei as well as the properties of symmetric and asymmetric nuclear matter. In present analysis we have adopted the well-known DDME2 parameter set \cite{lala05} to study the fusion barrier characteristics and also compared the results with the ones obtained using non-linear NL1 \cite{rein86}, NL3 \cite{lala97}, and TM1 \cite{bodmer91} parameter sets.
\begin{figure}
  \centering
  \includegraphics[scale=0.35]{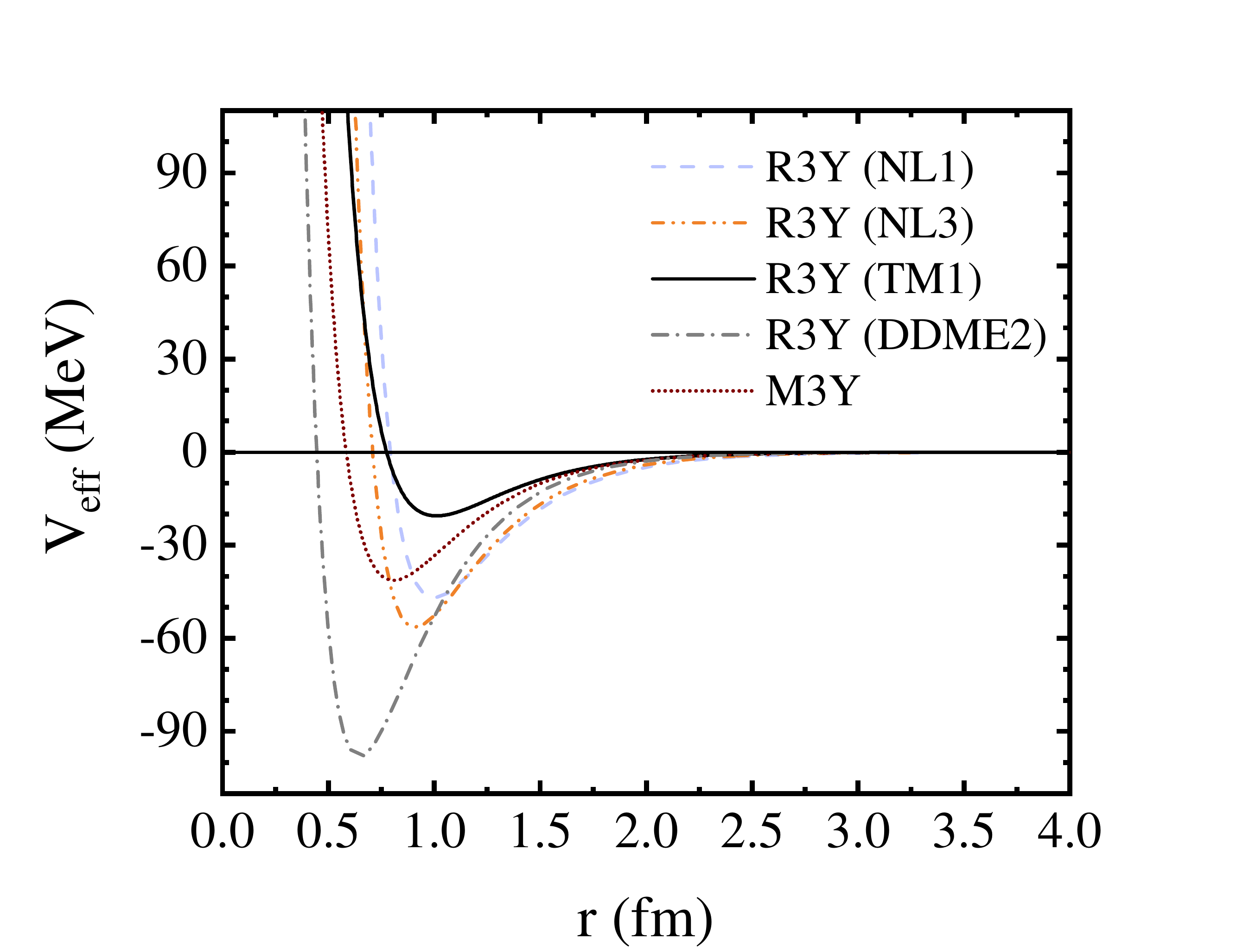}
  \caption{(Color online) The relativistic R3Y NN potential for NL1 (dashed light blue line), NL3 (dash double dotted orange line), TM1 (solid black line) and DDME2 (dash dotted grey line) parameter sets are compared with the phenomenological M3Y NN-interaction potential (dotted wine line). The R3Y NN potential for DDME2 is plotted at saturation density ($\rho_{sat}=0.152$ $fm^{-3}$ for DDME2 \cite{lala05}).}
  \label{fig1}
\end{figure}
\subsection{Medium-Dependent Relativistic R3Y Potential} 
The nucleon-nucleon interaction potential analogous to M3Y potential (see Eq. (\ref{m3y})) has also been derived by solving the mean-field equations in Eq. (\ref{field}) within the limit of one meson exchange \cite{sing12,sahu14,lahi16,bhuy20}. The relativistic NN-potential entitled as R3Y potential \cite{sing12,sahu14,lahi16,bhuy20} can be written as,
\begin{eqnarray}
V_{eff}^{R3Y}(r,\rho)&=&\frac{[g_{\omega}(\rho)]^{2}}{4{\pi}}\frac{e^{-m_{\omega}r}}{r} +\frac{[g_{\rho}(\rho)]^{2}}{4{\pi}}\frac{e^{-m_{\rho}r}}{r}\nonumber \\
&&
-\frac{[g_{\sigma}(\rho)]^{2}}{4{\pi}}\frac{e^{-m_{\sigma}r}}{r} +\frac{g_{2}^{2}}{4{\pi}} r e^{-2m_{\sigma}r}\nonumber \\
&& +\frac{g_{3}^{2}}{4{\pi}}\frac{e^{-3m_{\sigma}r}}{r}-\frac{\xi^2_3}{4\pi}\frac{e^{-3m_\omega r}}{r}+J_{00}(E)\delta(r). \nonumber \\
\label{r3y}
\end{eqnarray}
It is worth mentioning here that for the case of medium-independent R3Y NN potential obtained for NL1, NL3 and TM1 parameter sets i.e., the nucleon-meson couplings ($g_\sigma$, $g_\omega$ and $g_\rho$) appearing in Eq. (\ref{r3y}) are independent of the density. However, in case of medium-dependent DDR3Y NN potential for DDME2 parameter set, the $g_\sigma$, $g_\omega$ and $g_\rho$ are density-dependent [given in Eqs. (\ref{dd1}-\ref{dd3})] and also the non-linear self interaction constants ($g_2$, $g_3$, $\xi_3$) are zero for DDR3Y. The density ($\rho$) entering in Eqs. (\ref{dd1}-\ref{r3y}) is obtained within the relaxed density approximation (RDA) \cite{denisov13,denisov15} at the midpoint of the inter-nucleon separation distance and can be written as, 
\begin{eqnarray}
\rho\bigg(\frac{\vec{r}}{2}\bigg)&=&\rho_p(\vec{r}_p+\frac{\vec{r}}{2})-\frac{2\rho_{N_{p}}(\vec{r}_p+\frac{\vec{r}}{2})\rho_{N_{t}}(\vec{r}_t-\frac{\vec{r}}{2})}{\rho_{1_N}+\rho_{2_N}}\nonumber \\
&&+\rho_t(\vec{r}_t-\frac{\vec{r}}{2}) -\frac{2\rho_{P_{p}}(\vec{r}_p+\frac{\vec{r}}{2})\rho_{P_{t}}(\vec{r}_t-\frac{\vec{r}}{2})}{\rho_{1_P}+\rho_{2_P}}.
\label{rda}
\end{eqnarray}
Here, $\rho_{N_p}$ ($\rho_{P_p}$) and $\rho_{N_t}$ ($\rho_{P_t}$) are the neutron (proton) densities of projectile and target nuclei, respectively. Also, $\rho_{1(2)_{N}}=\frac{N_{p(t)}}{A_{p(t)}}\rho_{sat}$ and $\rho_{1(2)_{P}}=\frac{Z_{p(t)}}{A_{p(t)}}\rho_{sat}$, with $N_{p(t)}$ and $A_{p(t)}$ being the neutron and mass numbers of projectile (target) nuclei, respectively. More details about the validity of this RDA in obtaining the DDR3Y NN potential can be found in one of our recent works \cite{rana22}, where a comprehensive analysis of the fusion cross-section obtained using density-dependent M3Y \cite{khoa94,satc79,kobo82} and R3Y NN potentials are accomplished. In the present analysis, a systematic study of fusion barrier characteristics obtained using different RMF density distributions, relativistic R3Y, DDR3Y and non-relativistic M3Y NN potentials is done for 24 isospin asymmetric reaction systems forming heavy and superheavy nuclei. In open-shell nuclei, the pairing plays a significant role in describing their structure properties, including the density distributions. In the present study, we have considered the nuclei near the $\beta$-stable region of the nuclear chart, so we have considered the simple BCS pairing to take care of the pairing correlations \cite{zeng83,moli97,bhuy18,bhuy20}. Also, a blocking procedure is used to treat the odd-mass number nuclei \cite{bhuy18,bhuy20,doba84,madl88}.

The relativistic R3Y NN potential for different parameter sets and the analogous M3Y potential are shown in Fig. \ref{fig1}. The R3Y NN potential for DDME2 is plotted here for the coupling constants in Eq. (\ref{r3y}) at the saturation density, $\rho_{sat}=0.152$ $fm^{-3}$ \cite{lala05}. It can be observed from Fig. \ref{fig1} that the curves for R3Y NN-interaction potential for NL1, NL3, TM1, and DDME2 parameter sets show similar trends as the M3Y NN potential. The R3Y NN potential for the DDME2 parameter set at saturation density is observed to show the deepest pocket. However, the actual DDR3Y NN potential for the DDME2 parameter used to obtain the nuclear potential within the double-folding approach is density-dependent and is not exactly similar to one plotted at $\rho_{sat}=0.152$ $fm^{-3}$ in Fig. \ref{fig1}. A more detailed inspection shows that the R3Y NN potential for NL3 and NL1 parameter sets shows a slightly deep pocket compared to M3Y NN potential. However, the R3Y potential for the TM1 parameter shows a lightly shallow pocket compared to the M3Y potential, which can be connected with the self-coupling non-linear term ($\xi_3$) in the $\omega$-field. More details can be found in Refs. \cite{sing12,sahu14,lahi16,bhuy20}. The total potential, i.e., the barrier characteristics in Eq. (\ref{vtot}), is used further to obtain the fusion and/or capture cross-section using the $\ell$-summed Wong model. \\
\subsection{$\ell$-summed Wong model}
The fusion and/or capture cross-section for two colliding nuclei is given in terms of $\ell$ -partial wave by \cite{kuma09,bhuy18,bhuy20,wong73}
 \begin{eqnarray}
\sigma(E_{c.m.})=\frac{\pi}{k^{2}} \sum_{\ell=0}^{\ell_{max}}(2\ell+1)P_{\ell}(E_{c.m}).
\label{crs}
\end{eqnarray}
Here, $E_{c.m.}$ is the center of mass energy of target-projectile system and $k=\sqrt{\frac{2 \mu E_{c.m.}}{\hbar^{2}}}$. $P_{\ell}$ is called the penetration probability which describe the transmission through the barrier given in Eq. (\ref{vtot}). Using the Hill-Wheeler \cite{hill53} approximation, $P_{\ell}$ can be written in terms of barrier height ($V_B^\ell$) and curvature ($\hbar \omega_{\ell}$) as,
\begin{eqnarray}
P_{\ell}=\Bigg[1+exp\bigg(\frac{2 \pi (V_{B}^{\ell}-E_{c.m.})}{\hbar \omega_{\ell}}\bigg)\Bigg]^{-1}. 
\end{eqnarray}
Eq. (\ref{crs}) describes the fusion and/or capture cross-section of two interacting nuclei in terms of summation over $\ell$ partial waves. C. Y. Wong \cite{wong73}  replaced this summation by integration using the following approximation: (i) $\hbar \omega_{\ell}\approx \hbar \omega_{0}$, and (ii) $V_{B}^{\ell}\approx V_{B}^{0}+\frac{\hbar^2\ell(\ell+1)}{2\mu {R_{B}^{0}}^2}$, assuming $R_{B}^{\ell}\approx R_{B}^{0}$. These approximations lead to simple formula to estimate the fusion and/or capture cross-section in terms of barrier characteristics. This simplified Wong formula \cite{wong73} can be written as,
\begin{eqnarray}
\sigma(E_{c.m.}) &=& \frac{{R_{B}^{0}}^{2}\hbar\omega_{0}}{2E_{c.m.}}  
ln\bigg[1 + exp\bigg(\frac{2\pi}{\hbar\omega_{0}}(E_{c.m.}-V_{B}^{0})\bigg)\bigg].  \nonumber  \\
\label{wng}
\end{eqnarray}
However, using only $\ell = 0$ barrier and ignoring the modifications entering due to $\ell$- dependence of the potential cause the overestimation of fusion and/or capture cross-section by Wong formula at above barrier energies. Gupta and collaborators gave the solution of this problem \cite{kuma09,bhuy18,bhuy20} by using the more precise $\ell$-summed formula given in Eq. (\ref{crs}).
\begin{figure*}
    \centering
    \includegraphics[scale=0.65]{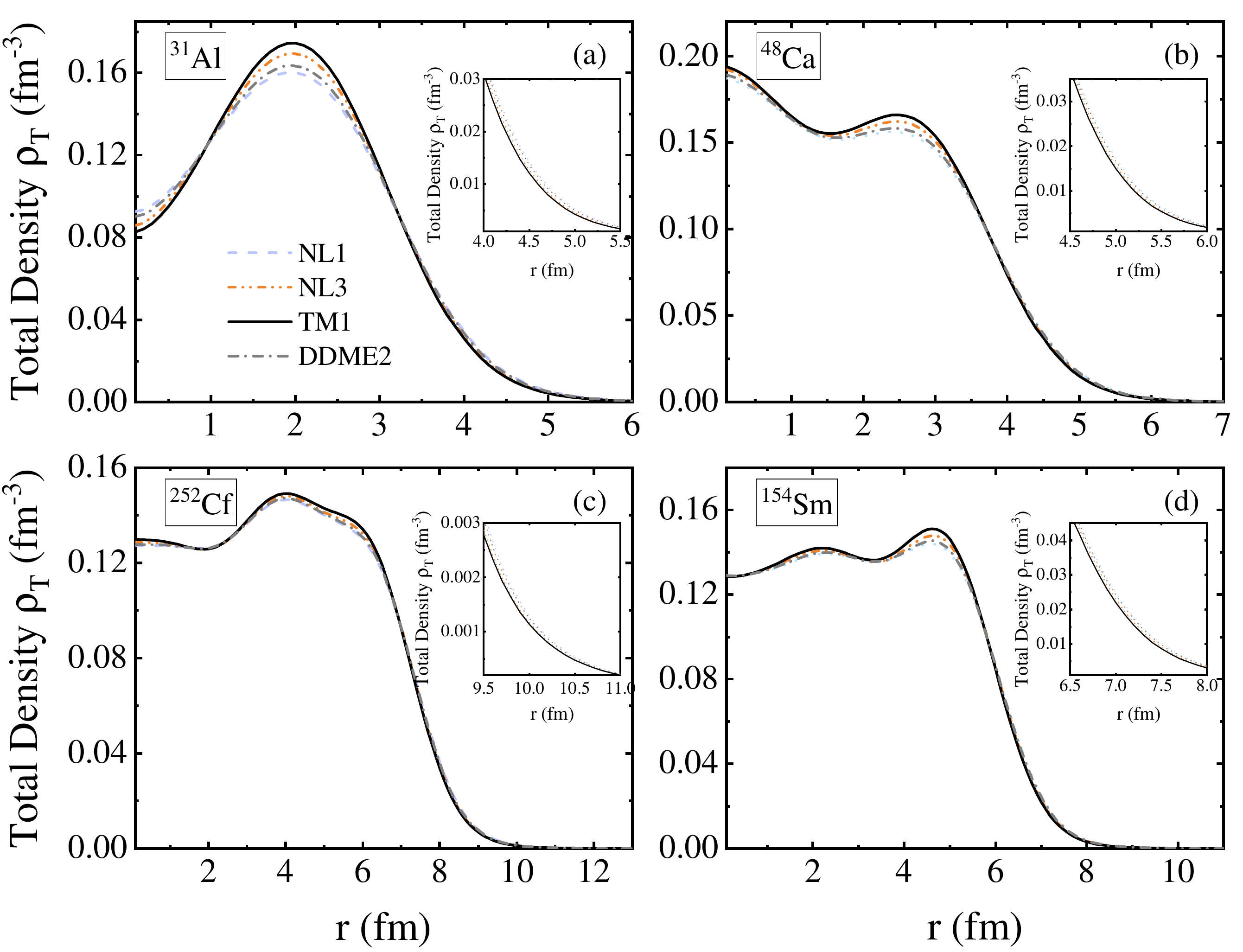}
    \caption{(Color online) The variation of total density distributions as a function of nuclear radius r (fm) for (a) $^{31}$Al, (b) $^{48}$Ca, (c) $^{252}$Cf and (d) $^{154}$Sm  nuclei calculated using RMF with NL1 (light blue), NL3 (orange), TM1 (black), and DDME2 (grey) parameter sets. Insets shows the magnified view of tail region of the densities. See text for details.}
    \label{fig2}
\end{figure*}

\section{CALCULATIONS AND DISCUSSION}
\label{results} 
In this section, the fusion and/or capture cross-section of heavy-ion reactions is studied within the $\ell-$summed Wong model supplemented with self-consistent relativistic mean-field and relativistic-Hartree-Bogoliubov formalism. As we know, the value of nuclear matter observable and the structural properties of finite nuclei obtained from RMF and RHB formalism depend upon the force parameter set. Parallel to this, the dependence of fusion characteristics upon these RMF parameter sets can be anticipated. In the present analysis, we systematically study the nuclear density distributions and the effective nucleon-nucleon interaction potential dependence on the different parameter sets and, consequently, on the fusion barrier characteristics. The nuclear density distributions for all the interacting nuclei (targets and projectiles) are obtained within the relativistic mean-field formalism for three force parameter sets NL1, NL3 and TM1, and within the Relativistic-Hartree-Bogoliubov approach for the DDME2 parameter set. We have considered three types of effective nucleon-nucleon interactions: (i) The widely used non-relativistic M3Y potential given by Eq. (\ref{m3y}). (ii) Density independent relativistic R3Y NN potential described in terms of masses and coupling constants of the mesons ($\sigma$-, $\omega$- and $\rho$- mesons). The medium independent relativistic R3Y potential is obtained for three relativistic parameter sets, i.e., NL1, NL3 and TM1. (iii) The medium dependence of the R3Y NN potential is also taken into account via density-dependent meson-nucleon coupling terms obtained from the DDME2 parameter set.

From Fig. \ref{fig1}, it can be noted that the NN potentials of M3Y and R3Y for NL1, NL3, TM1, and DDME2 sets show a similar trend but with different depths. A systematic study is carried out to study the effect of these NN interactions on the nuclear potential ($V_n$) calculated within the double folding approach. As mentioned above, the R3Y NN potential for DDME2 parameter is plotted at saturation density ($\rho_{sat}=0.152$ $fm^{-3}$) in Fig. \ref{fig1} whereas actual density-dependent R3Y (DDR3Y) obtained within the RHB approach is used for calculating the nuclear potential and fusion and/or capture cross-section. The calculations are done in three steps: i) Folding of nuclear density distributions with the medium independent relativistic R3Y potential for three parameters sets NL1, NL3 and TM1, and DDR3Y NN potential for DDME2 parameter set to obtain the nuclear potential. The nuclear density distributions are also folded with the phenomenological M3Y potential for the sake of comparison. ii) In the second step, the R3Y potential is fixed for one parameter set and is folded with the density distributions obtained with the considered four parameter sets. iii) In the last step, the density is fixed for one parameter set and is folded with the R3Y NN potential obtained for NL1, NL3, TM1, and DDME2 parameter sets. \\ \\
{\bf Nuclear Density Distributions:} The RMF formalism successfully describes the bulk properties such as the binding energy, quadrupole deformations, nuclear density distributions, etc., throughout the nuclear chart. The total density (sum of the proton and neutron number densities, i.e., $\rho_T= \rho_P+ \rho_N$) as a function of nuclear radius (r) is plotted in Fig. \ref{fig2} for the representative case of  (a) odd mass $^{31}$Al, (b) even mass light $^{48}$Ca, (c) heavy $^{252}$Cf and (d) intermediate $^{154}$Sm  nuclei. These density distributions are obtained by solving the RMF equations for the NL1 (dashed light blue line), NL3 (dash double dotted orange line), TM1 (solid black line) parameter sets and within the RHB approach for DDME2 (dash dotted grey line) parameter set. It can be observed from Fig. \ref{fig2}(a) that the density of odd mass light $^{31}$Al nucleus shows a peak in the central region whereas the density of $^{48}$Ca nucleus shows a downturn in the central region which is caused by the combined effects of Coulomb repulsion and the nuclear shell structure \cite{rein02,afan05,chu10}. Moreover, the density distributions of intermediate and heavy mass target nuclei show a comparatively flattened curve in the central region that falls rapidly in the surface region. The figure shows that the NL1 and TM1 have respectively, the lowest and highest magnitudes of central density for all the nuclei under study. In the case of the heavy-ion fusion reactions, the density at the tail/surface region only plays the most crucial role in the fusion process \cite{raj07}. The insets in Fig \ref{fig2} show the magnified view of the tail region of the densities. A slight difference is observed among the NL1, NL3, TM1 and DDME2 density distributions at the surface/tail region of all the interacting nuclei. A systematic and quantitative study is carried out in the upcoming subsections to analyze the effects of this slight difference on the fusion characteristics.
\subsection{Folding RMF densities with R3Y, DDR3Y and M3Y NN potentials:}
In the first analysis, a comparison is made between the widely adopted non-relativistic M3Y, density-independent relativistic R3Y and the density-dependent relativistic R3Y (DDR3Y) NN potentials within the double folding approach. The RMF nuclear density distributions obtained for non-linear NL1, NL3, TM1 and density-dependent DDME2 parameter sets (see Fig. \ref{fig2}) are folded with the M3Y, R3Y and DDR3Y NN potentials to obtain the nuclear optical potential. In our earlier work, the density-independent M3Y and R3Y nucleon-nucleon potentials have been used to study the fusion hindrance phenomenon in a few Ni-based reactions \cite{bhuy18} and to study the cross-section for the synthesis of heavy and superheavy nuclei \cite{bhuy20,rana21}. In \cite{bhuy18,bhuy20}, the fusion and/or capture cross-section obtained using non-relativistic M3Y NN potential is compared with relativistic R3Y NN potential for NL3$^*$ parameter set only. A more systematic study with the inclusion of different non-linear RMF parameter sets as well as the explicit medium dependence of R3Y NN potential is carried out in the present analysis to investigate the effect of different nucleon-nucleon potentials and the nuclear density distributions on the fusion and/or capture cross-section of 24 different exotic reaction systems. The reaction system involves the projectile with a higher N/Z ratio, synthesizing neutron-rich heavy and superheavy nuclei. We have considered 17 target-projectile combinations for the synthesis of different isotopes of SHN with Z=120. The fusion barrier characteristics (barrier height, barrier position, frequency, etc.) are obtained within the double folding approach for both M3Y and R3Y NN potential. Then the fusion and/or capture cross-section is calculated from the well-known $\ell-$summed Wong model.


\begin{figure*}
\centering
\includegraphics[scale=0.6]{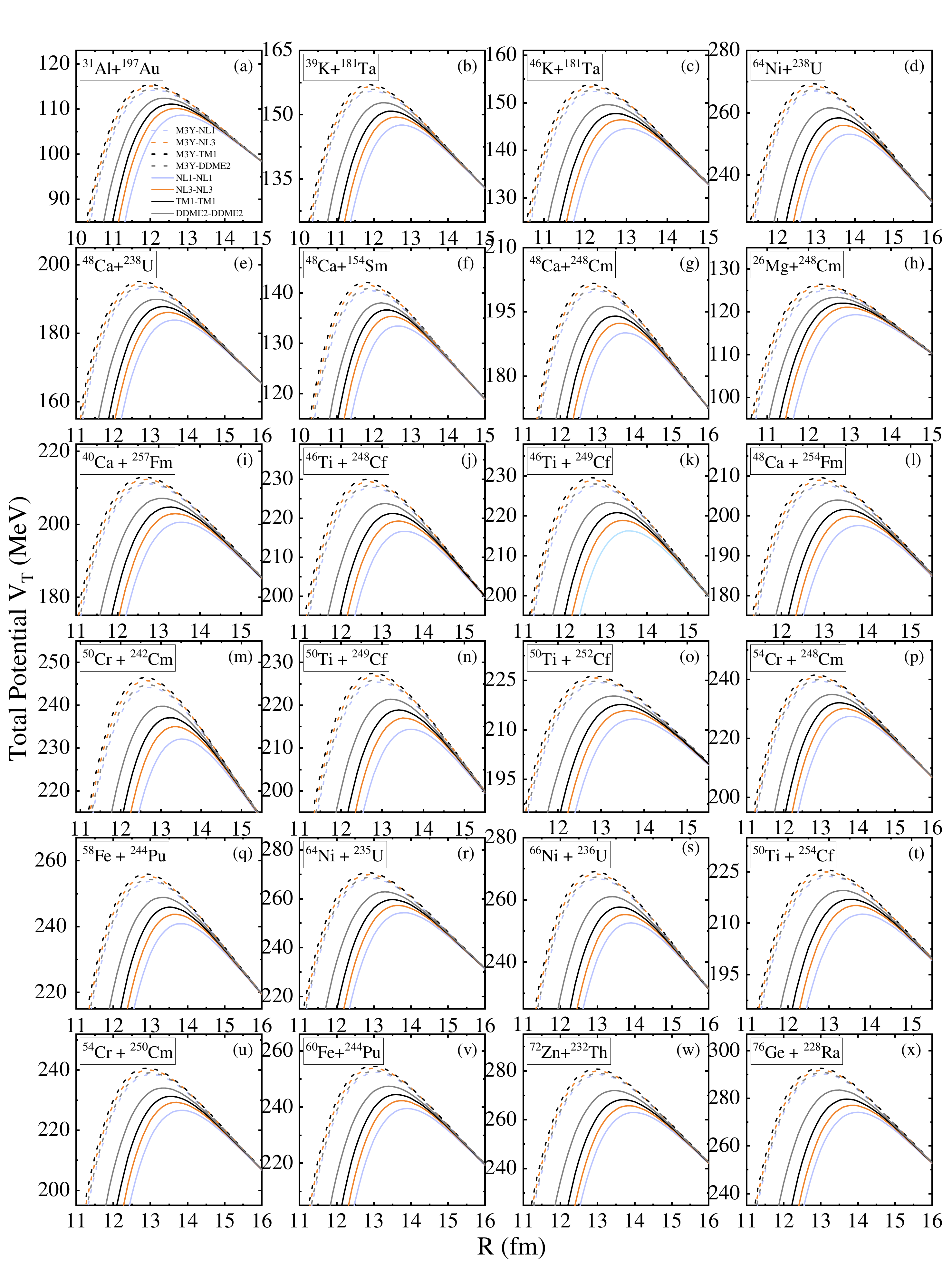}
\caption{(Color online) The total interaction potential $V_T$ (MeV) at $\ell =0\hbar$ as a function of radial separation R for 24 reaction systems under study calculated using the M3Y (dashed lines), R3Y and DDR3Y (solid lines) NN potentials. The different colours are for parameter sets as labeled in the figure. See text for details.}
\label{fig3}
\end{figure*}

\noindent
{\bf Total Interaction Potential:} As discussed above, the interaction potential at $\ell =0 \hbar$ (sum of nuclear and Coulomb potential) formed between the target and projectile nuclei plays the most crucial role in determining the fusion characteristics of the system. We have calculated the nuclear interaction potential from nuclear density distributions integrated over M3Y, R3Y and DDR3Y effective NN interactions. The total interaction potential at $\ell =0 \hbar$ [$V_T(R)=V_n(R)+V_C(R)$] is obtained for the even-even $^{48}$Ca+$^{154}$Sm, $^{48}$Ca+$^{238}$U, $^{48}$Ca+$^{248}$Cm, $^{26}$Mg+$^{248}$Cm; even-odd $^{46}$K+$^{181}$Ta; and odd-odd $^{31}$Al+$^{197}$Au and $^{39}$K+$^{181}$Ta systems and also for 17  systems for the synthesis of SHN Z=120 $^{40}$Ca + $^{257}$Fm, $^{48}$Ca + $^{254}$Fm, $^{46}$Ti + $^{248}$Cf, $^{46}$Ti + $^{249}$Cf, $^{50}$Ti + $^{249}$Cf, $^{50}$Ti + $^{252}$Cf, $^{50}$Cr + $^{242}$Cm, $^{54}$Cr + $^{248}$Cm, $^{58}$Fe + $^{244}$Pu, $^{64}$Ni+$^{238}$U, $^{64}$Ni + $^{235}$U, $^{66}$Ni + $^{236}$U, $^{50}$Ti + $^{254}$Cf, $^{54}$Cr + $^{250}$Cm, $^{60}$Fe + $^{244}$Pu, $^{72}$Zn + $^{232}$Th and $^{76}$Ge + $^{228}$Ra. Fig. \ref{fig3} shows the barrier region of the total interaction potential (MeV) at $\ell =0\hbar$ as a function of the radial separation $R$ ($fm$) for all the 24 reaction systems. The dashed lines signify that the phenomenological M3Y NN potential integrated over the nuclear densities obtained for NL1 (light blue), NL3 (orange), TM1 (black), and DDME2 (grey) parameter sets. However, the solid lines signify the relativistic R3Y and DDR3Y potential along with the mean-field densities are used to obtain the nuclear potential within the double folding approach. 
\begin{figure*}
    \centering
    \includegraphics[scale=0.335]{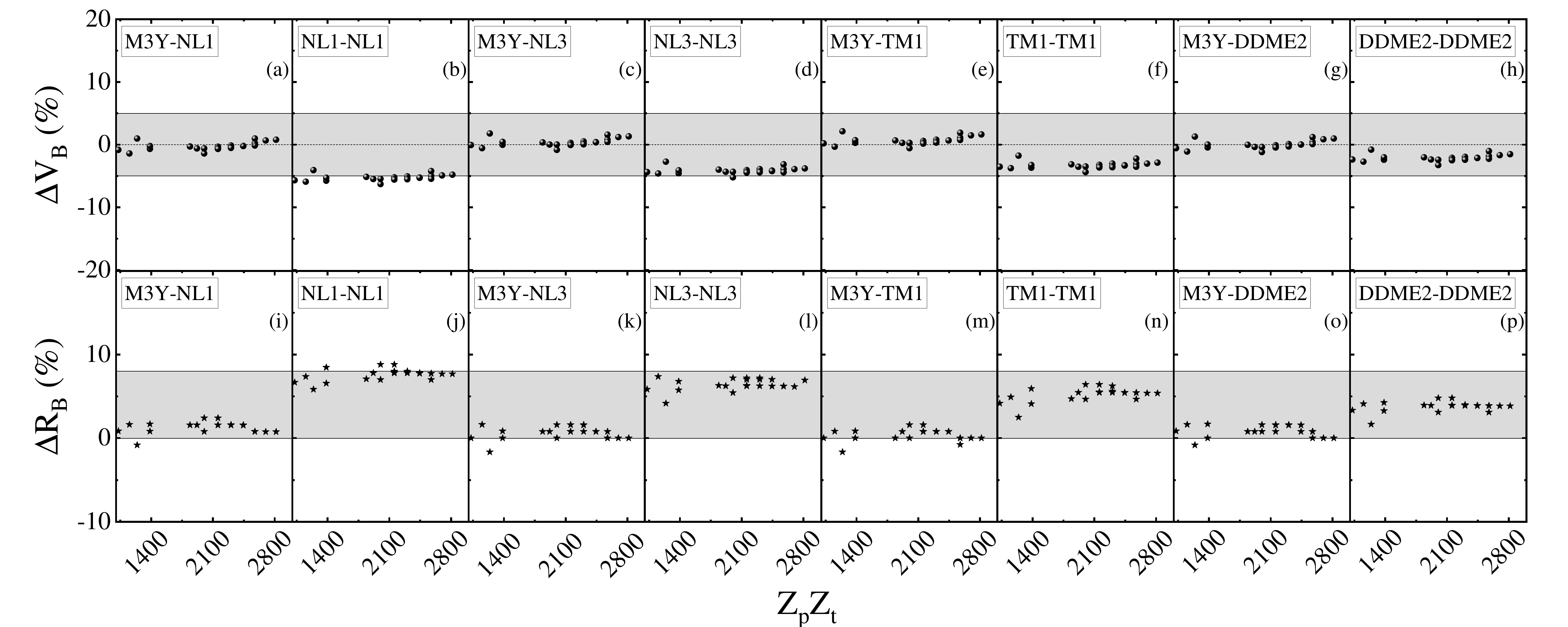}
    \caption{The percentage (\%) change in barrier height (upper panel) and barrier position (lower panel) as a function of charges $Z_pZ_t$ for all the 24 reaction systems under study. See text for details.}
    \label{fig4}
\end{figure*}

From Fig. \ref{fig3}, one can notice that the M3Y NN potential gives a relatively higher barrier as compared to the relativistic R3Y and DDR3Y NN potentials for all the considered systems. This signifies that the microscopic R3Y effective NN potential given in terms of the meson masses and their coupling constants gives comparatively more attractive interaction potential than the M3Y NN potential, described as the sum of three Yukawa terms. Comparing the barrier heights for different density distributions (NL1, NL3, TM1, and DDME2) folded with M3Y potential, it is observed that TM1 and NL1 sets give the highest and lowest barrier heights, respectively. For the case of M3Y NN potential, the NL3 and TM1 densities are observed to give a higher barrier than the DDME2 parameter set whereas for the case R3Y NN potential the DDME2 set gives the highest barrier. This is because the R3Y NN potential for the DDME2 parameter is density-dependent. The relaxed density approximation (RDA) is used here to include the density dependence of microscopic R3Y NN potential in terms of nucleon-meson couplings. Thus, the barrier height is observed to be increased with the inclusion of the in-medium effects of the R3Y NN potential. 

\noindent
The Bass potential \cite{bass73,bass74,bass77} is the simpler and well-known form of nuclear potential. Here, we have studied the variation in fusion barrier characteristics i.e., barrier height $V_B$ and position $R_B$ obtained from M3Y, R3Y and DDR3Y NN potentials with respect to (w.r.t.) those obtained from Bass potential. Fig. \ref{fig4} shows the percentage change in barrier height (upper panel) and barrier position (lower panel) as a function of charges  $Z_pZ_t$ for all the 24 systems. Here,  $Z_p$ and $Z_t$ are atomic numbers for projectile and target nuclei, respectively. M3Y-NL1, M3Y-NL3, M3Y-TM1, and M3Y-DDME2 signify that the nuclear density distributions obtained for non-linear NL1, NL3, TM1 and density-dependent DDME2 parameter sets, respectively, along with the M3Y NN potential, are used within the double folding procedure to calculate the nuclear potential. Similarly, NL1-NL1, NL3-NL3, and TM1-TM1 signify that the RMF density distributions and relativistic R3Y NN potential are used within the double folding approach to obtain the nuclear potential. Also, DDME2-DDME2 signify that the DDR3Y NN potential and density distributions obtained within the RHB approach for the DDME2 parameter set are used to calculate the nuclear potential. The Coulomb potential is added then to this nuclear potential potential and the barrier characteristics ($V_B$, $R_B$) are obtained using Eq. (\ref{vb1}) and (\ref{vb2}). It can be noted from Fig. \ref{fig4} that the nuclear potential calculated for M3Y NN potential shows $\le2\%$ change in barrier height and $\le1\%$ change in barrier position for the considered reactions w.r.t. the Bass potential. However, in the case of nuclear potential obtained for R3Y and DDR3Y NN potentials, the barrier height decreases by up to $\approx5\%$ and the barrier position shifts by up to $\approx8\%$ towards the higher separation distance w.r.t. the Bass potential. Moreover, this percentage change in barrier characteristics is minimum for the nuclear density distributions obtained for the TM1 parameter set and is maximum for those for the NL1 parameter set. This shows that the inclusion of vector meson self-coupling term ($\propto \omega^4$) in RMF Lagrangian results in the stronger repulsive core of NN interaction potential. The characteristics of the fusion barrier have a direct impact on the fusion cross-section of the reaction systems. The higher the barrier height value calculated for a system, the lower will be its cross-section. The effect of different RMF densities and the NN interaction potential on the fusion and/or capture cross-section is studied using the well known $\ell-$summed model in the following subsection.
\begin{figure*}
\centering
\includegraphics[scale=.6]{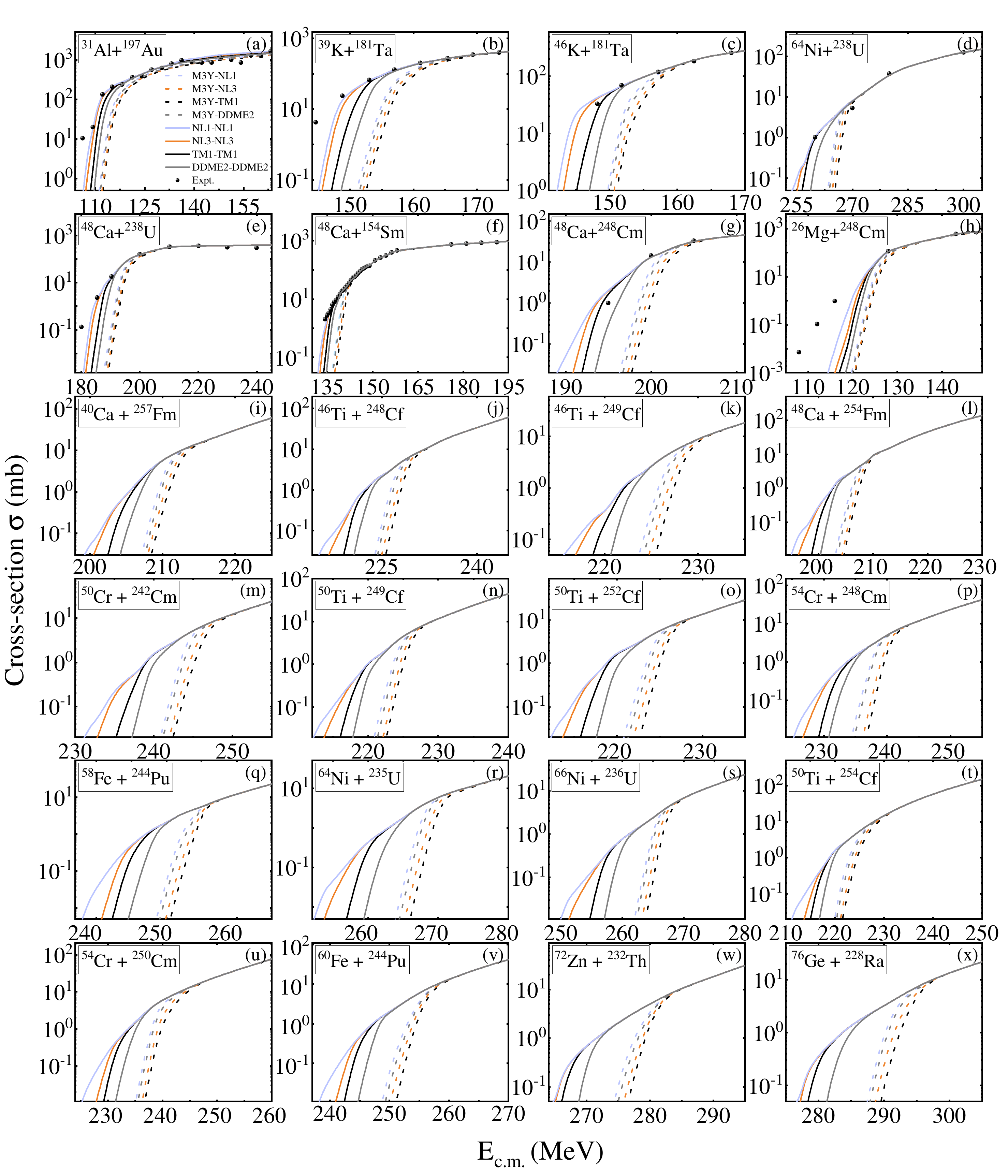}
\caption{(Color online) The cross-section $\sigma$ ($mb$) for all target projectile combinations considered in the present study by using M3Y (dashed lines) and R3Y (solid lines) interaction using NL1 (light blue), NL3 (orange), TM1 (black) and DDME2 (grey) parameter sets. The solid black circle indicates the experimental data \cite{wata01,knya07,itki04,prok05,kozu10,wakhle18}. See text for details.}
\label{fig5}   
\end{figure*}

\noindent 
{\bf fusion and/or capture Cross-Section:} The characteristics of total interaction potential (barrier height, position, and frequency) are further used to estimate the fusion probability and cross-section. We have calculated the fusion and/or capture cross-section for all the 24 reaction systems within the well-known $\ell-$summed Wong model described in detail in the previous section. Fig. \ref{fig5} shows the cross-section $\sigma$ (mb) as function of center of mass energy $E_{c. m.}$ (MeV) for all the target-projectile systems including the even-even $^{48}$Ca+$^{154}$Sm, $^{48}$Ca+$^{238}$U, $^{48}$Ca+$^{248}$Cm, $^{64}$Ni+$^{238}$U, $^{26}$Mg+$^{248}$Cm; even-odd $^{46}$K+$^{181}$Ta; and odd-odd $^{31}$Al+$^{197}$Au and $^{39}$K+$^{181}$Ta systems. We have also considered here 17 different target-projectile combinations which lead to the synthesis of SHN Z=120 $^{40}$Ca + $^{257}$Fm, $^{48}$Ca + $^{254}$Fm, $^{46}$Ti + $^{248}$Cf, $^{46}$Ti + $^{249}$Cf, $^{50}$Ti + $^{249}$Cf, $^{50}$Ti + $^{252}$Cf, $^{50}$Cr + $^{242}$Cm, $^{54}$Cr + $^{248}$Cm, $^{58}$Fe + $^{244}$Pu, $^{64}$Ni+$^{238}$U, $^{64}$Ni + $^{235}$U, $^{66}$Ni + $^{236}$U, $^{50}$Ti + $^{254}$Cf, $^{54}$Cr + $^{250}$Cm, $^{60}$Fe + $^{244}$Pu, $^{72}$Zn + $^{232}$Th and $^{76}$Ge + $^{228}$Ra. The calculated fusion and/or capture cross-section is also compared with the available experimental data \cite{wata01,knya07,itki04,prok05,kozu10,wakhle18}. The $\ell_{max}$ values are calculated using the sharp cut-off model \cite{beck81} for the reaction systems with experimental data. Since the experimental fusion and/or capture cross-section is not available for all the systems leading to the formation of SHN Z=120  except $^{64}$Ni+$^{238}$U reaction \cite{kozu10}, so sharp cut-off model is not applicable for them. To extract the $\ell_{max}$ values for these system we have used the polynomial between $E_{c.m.}/V_B$ and $\ell_{max}$ values constructed  using $^{64}$Ni+$^{238}$U data in our earlier work \cite {rana21}. The dashed lines in Fig. \ref{fig5} show the cross-section estimated by employing the nuclear potential calculated by folding the M3Y NN interaction over the nuclear density distribution obtained for NL1 (light blue), NL3 (orange), TM1 (black) and DDME2 (grey) parameter sets. The solid lines in Fig. \ref{fig5} signify that the nuclear potential calculated by folding the relativistic R3Y and DDR3Y NN potentials along with the spherical densities obtained for NL1 (light blue), NL3 (orange), TM1 (black) and DDME2 (grey) parameter sets is employed to calculate the cross-section.\\ \\
It can be observed from Fig. \ref{fig5} that the R3Y and DDR3Y NN interaction potentials give much better overlap with the experimental data as compared to the M3Y NN interaction for all the reaction systems. Comparing the cross-section from different relativistic force parameter sets, we find that the DDME2 parameter set gives the lowest, whereas the NL1 set gives the highest cross-section value for a given system. However, this difference is more evident at below and around the fusion barrier centre-of-mass energies. At above barrier energies, the cross-sections from all the parameter set almost overlap. The reason for this behaviour is that the structure effects get diminished at above barrier energies, and the angular momentum part of total potential dominates \cite{bhuy18}. Comparison of the results for different relativistic parameter sets with the experimental data shows that the NL1 parameter set is superior to the NL3, TM1 and DDME2 parameter sets. Out of NL3, TM1 and DDME2 sets, the parameter set NL3 is observed to fit the experimental data better. This is because the TM1 parameter set, which includes the self-coupling terms of the $\omega$-mesons, gives comparatively repulsive NN interaction, underestimating the fusion and/or capture cross-section. However, for three reactions namely $^{46}$K+$^{181}$Ta (Fig. \ref{fig5}(b)), $^{64}$Ni+$^{238}$U (Fig. \ref{fig5}(d)) and $^{48}$Ca+$^{248}$Cm (Fig. \ref{fig5}(g)), TM1 give better overlap than the other parameter sets. Moreover, the DDME2 density folded with M3Y is observed to give a higher cross-section than NL3 and TM1 densities. On contrary, the DDME2 density folded with DDR3Y NN potential gives a lower cross-section than the R3Y NN potential folded with NL1, NL3 and TM1 densities. This indicates that the inclusion of density dependence of microscopic R3Y NN potential in terms of the DDME2 parameter set decreases the cross-section. In the case of the system $^{26}$Mg+$^{248}$Cm (Fig. \ref{fig5}(h)), both the M3Y as well as R3Y NN potentials underestimate the experimental cross-section at below-barrier energies. This deviation between the experimental and theoretically calculated cross-section is caused by the fusing nuclei's structural deformations, which are not considered in the present study. \\ \\
In the case of the reaction systems leading to the formation of different isotopes of SHN with Z=120, the experimental data is only available for the $^{64}$Ni+$^{238}$U and, as discussed above, shows a good overlap with the results obtained with the R3Y NN interaction for TM1 parameter sets. Among all the systems for Z=120, the difference between cross-sections obtained for NL1,  NL3, TM1, and DDME2 parameters sets is comparatively a little more prominent for $^{58}$Fe+$^{244}$Pu (see Fig. \ref{fig5}(v)) system. In the case of $^{26}$Mg+$^{248}$Cm system, the experimental data is available at the centre of mass energies far below its Bass barrier (at 126.833 MeV). For $^{48}$Ca+$^{248}$Cm system, the R3Y NN potential for TM1 gives a better fit to the experimental data as compared to other systems. From all these observations, a more systematic investigation of the effects of NN potential and RMF density distributions on the cross-section is carried only for these three reaction systems in the upcoming subsections. We have dropped out the M3Y NN potential for further investigation of fusion characteristics as it gives comparatively poor overlap with the experimental data. Also, since the NL1 and NL3 RMF parameter sets give comparatively better results than the TM1 parameter sets, we fix the NN R3Y potential and RMF density distribution for further second and third steps for these parameter sets to explore their effects on the fusion characteristics.
\begin{figure*}
\centering
\includegraphics[scale=0.76]{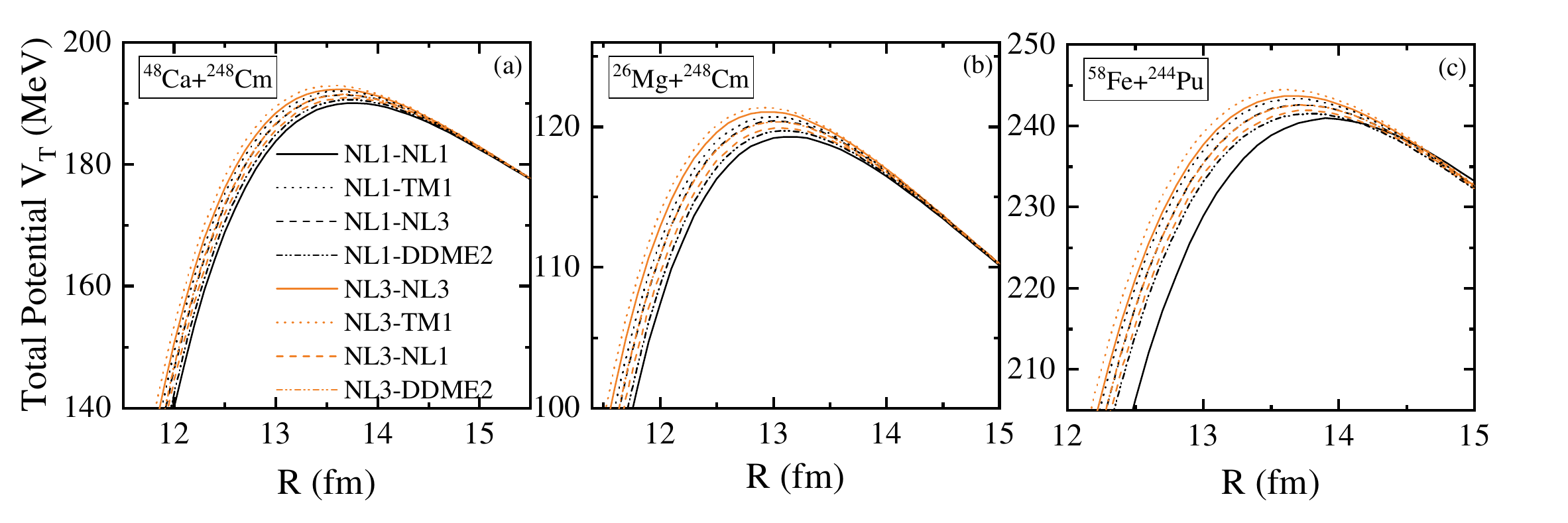}
\caption{(Color online) The variation of total potential $V_T$ (in MeV) as a function of radial distance $R$ (in $fm$) for (a) $^{48}$Ca+$^{248}$Cm, (b) $^{26}$Mg+$^{248}$Cm and (c) $^{58}$Fe+$^{244}$Pu calculated by fixing the NN potential for NL1 and NL3 parameter set. The NL3-TM1 signifies R3Y NN potential using NL3 parameter set and density distributions using TM1 parameters within folding procedures are used to obtain the nuclear potential. The same procedure is followed for other cases as labeled in the figure.} 
\label{fig6}
\end{figure*}
\begin{figure*}
\centering
\includegraphics[scale=0.90]{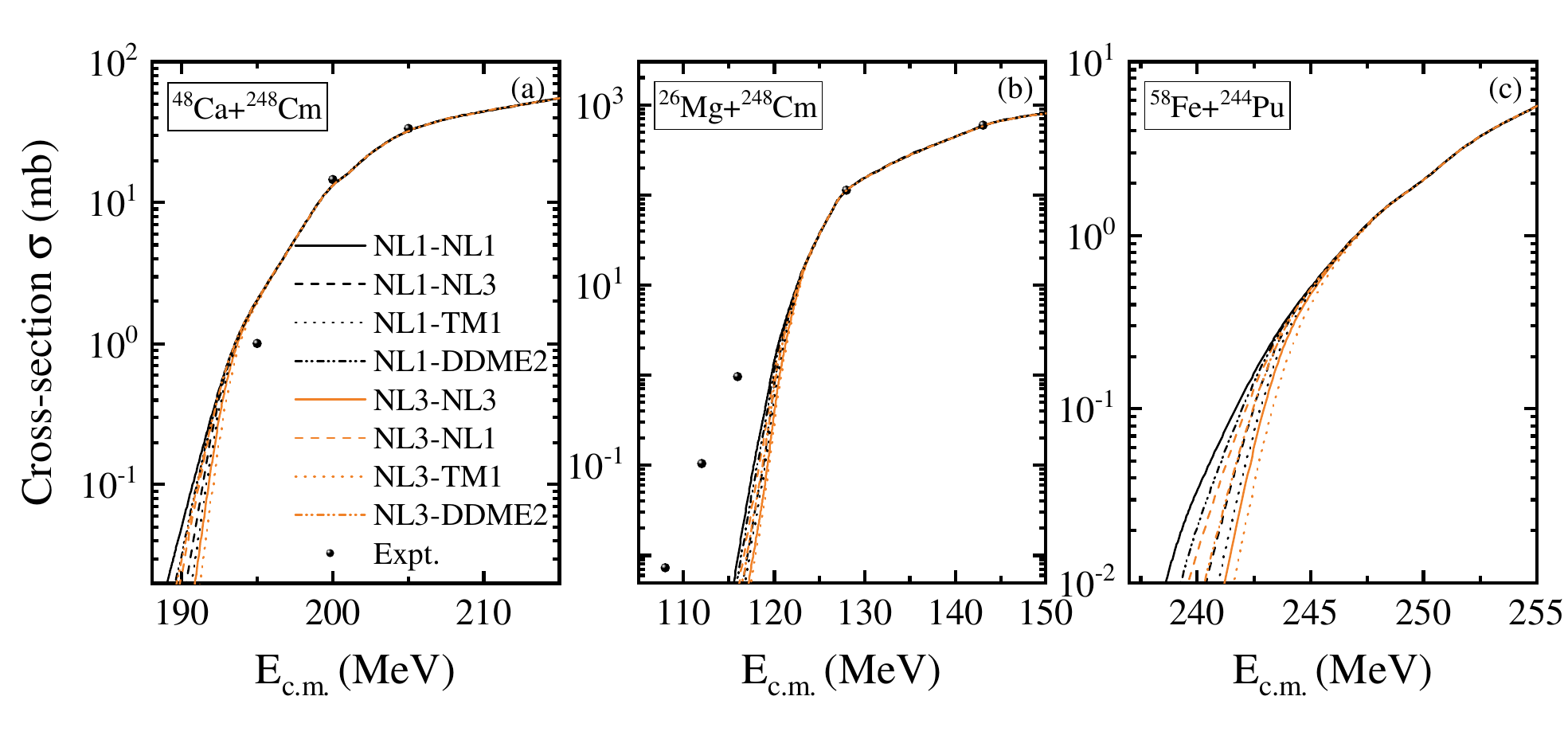}
\caption{(Color online) The cross-section $\sigma$ (mb) calculated with $\ell$-summed Wong model using R3Y NN potential for (a) $^{48}$Ca+$^{248}$Cm, (b) $^{26}$Mg+$^{248}$Cm and (c) $^{58}$Fe+$^{244}$Pu.  The NL3-TM1 signifies the R3Y NN potential using NL3 parameter set and density distributions using TM1 parameter set are used within folding procedures to obtain the nuclear potential. The same procedure is followed for other cases as labeled in the figure.}
\label{fig7}
\end{figure*}
\begin{table*}
\caption{\label{tab1} The barrier position $R_B$ (in $fm$) and barrier height $V_B$ (in MeV) for all the 24 considered reactions by fixing one effective NN interaction and varying the nuclear density distributions. The NL1-NL3 signifies that the R3Y NN potential obtained for NL1 parameter set is folded with RMF densities obtained for NL3 parameter set.}
\adjustbox{width=\textwidth}{
\begin{tabular}{lllllllllllllllll}
\hline
 \multicolumn{1}{c}{\multirow{2}{*}{System}} 
& \multicolumn{2}{c}{NL1-NL1} & \multicolumn{2}{c}{NL1-NL3}                               & \multicolumn{2}{c}{NL1-TM1}                               &
\multicolumn{2}{c}{NL1-DDME2}             &
\multicolumn{2}{c}{NL3-NL3}                               & \multicolumn{2}{c}{NL3-NL1}                               & \multicolumn{2}{c}{NL3-TM1}                            &
\multicolumn{2}{c}{NL3-DDME2}               \\ \cline{2-17}
\multicolumn{1}{c}{}                          & \multicolumn{1}{c}{$R_B$} & \multicolumn{1}{c}{$V_B$} & \multicolumn{1}{c}{$R_B$} & \multicolumn{1}{c}{$V_B$} & \multicolumn{1}{c}{$R_B$} & \multicolumn{1}{c}{$V_B$} & \multicolumn{1}{c}{$R_B$} & \multicolumn{1}{c}{$V_B$} & \multicolumn{1}{c}{$R_B$} & \multicolumn{1}{c}{$V_B$} &
\multicolumn{1}{c}{$R_B$} & \multicolumn{1}{c}{$V_B$} &
\multicolumn{1}{c}{$R_B$} & \multicolumn{1}{c}{$V_B$} &
\multicolumn{1}{c}{$R_B$} & \multicolumn{1}{c}{$V_B$} \\ \cline{1-17}
$^{31}$Al+$^{197}$Au  & 12.8 & 108.63 & 12.7 & 109.56 & 12.7 & 109.88 & 12.8 & 108.99 & 12.7 & 110.13 & 12.8 & 109.18 & 12.6 & 110.45 & 12.7 & 109.53 \\ 
 $^{39}$K+$^{181}$Ta & 12.8 & 147.56 & 12.7 & 148.67 & 12.8 & 148.94 & 12.7   & 147.99 & 12.6 & 149.44 & 12.7 & 148.31 & 12.6 & 149.90 &  12.7 & 148.72   \\ 
 $^{46}$K+$^{181}$Ta  & 13.0 & 144.60 & 12.9 & 145.73 & 12.9 & 146.15 &  13.0 & 145.03  & 12.9 & 146.44 & 13.0 & 145.28 & 12.8 & 146.85 & 12.9  & 145.71   \\ 
$^{64}$Ni+$^{238}$U  & 13.9 & 253.12 & 13.8 & 254.91 & 13.7 & 255.69 &  13.8 & 253.80  & 13.7 & 255.98 & 13.8 & 254.14 & 13.7 & 256.76 & 13.8  &  254.82 \\ 
 $^{48}$Ca+$^{238}$U  & 13.6 & 183.85 & 13.5 & 185.22 & 13.5 & 185.79 & 13.6  & 184.38  & 13.5 & 186.09 & 13.6 & 184.68 & 13.4 & 186.65 & 13.5  & 185.20  \\ 
 $^{48}$Ca+$^{154}$Sm & 12.7 & 133.44 & 12.5 & 134.64 & 12.5 & 135.15 &  12.6 & 133.90  & 12.5 & 135.34 & 12.6 & 134.11 & 12.4 & 135.85 &  12.5 & 134.56  \\ 
 $^{48}$Ca+$^{248}$Cm & 13.8 & 190.08 & 13.7 & 191.46 & 13.6 & 191.99 & 13.7  & 190.60  & 13.6 & 192.34 & 13.7 & 190.92 & 13.6 & 192.87 & 13.7  & 191.44  \\ 
 $^{26}$Mg+$^{248}$Cm  & 13.1 & 119.30 & 13.0 & 120.42 & 13.0 & 120.71 & 13.1
  & 119.73 & 13.1 & 120.96 & 13.1 & 119.93 & 12.9 & 121.37 & 13.0  & 120.35  \\ 
 $^{40}$Ca+$^{257}$Fm & 13.6 & 200.59 & 13.5 & 201.97 & 13.5 & 202.51 & 13.5
 & 201.12  & 13.4 & 202.93 & 13.5 & 201.53 & 13.4 & 203.51 & 13.5  & 202.05  \\ 
 $^{46}$Ti+$^{248}$Cf & 13.5 & 216.61 & 13.5 & 218.21 & 13.4 & 218.87 &  13.5
 & 217.23  & 13.4 & 219.25 & 13.5 & 217.62 & 13.4 & 219.90 &  13.5 &  218.21 \\ 
 $^{46}$Ti+$^{249}$Cf & 13.6 & 216.26 & 13.5 & 217.84 & 13.4 & 218.47 & 13.5
  &  216.85 & 13.4 & 218.86 & 13.5 & 217.24 & 13.4 & 219.51 &  13.5 & 217.85 \\ 
 $^{48}$Ca+$^{254}$Fm & 13.8 & 197.58 & 13.7 & 199.01 & 13.7 & 199.55 & 13.8
  &  198.10 & 13.6 & 199.90 & 13.7 & 198.43 & 13.6 & 200.48 & 13.7 & 199.00  \\ 
 $^{50}$Cr+$^{242}$Cm & 13.5 & 232.12 & 13.4 & 233.90 & 13.4 & 234.60 & 13.5
  & 232.79  & 13.4 & 235.00 & 13.5 & 233.17 & 13.3 & 235.72 & 13.4  & 233.87 \\ 
 $^{50}$Ti+$^{249}$ Cf & 13.7 & 214.40 & 13.6 & 215.99 & 13.6 & 216.58 &  13.7
 & 214.99  & 13.5 & 216.98 & 13.6 & 215.35 & 13.5 & 217.61 &  13.6 &  215.98 \\ 
 $^{50}$Ti+$^{252}$Cf & 13.8 & 213.32 & 13.7 & 214.88 & 13.6 & 215.46 &  13.7
 & 213.91  & 13.6 & 215.88 & 13.7 & 214.28 & 13.6 & 216.46 & 13.7  & 214.88  \\ 
 $^{54}$Cr+$^{248}$Cm & 13.8 & 227.41 & 13.7 & 229.06 & 13.7 & 229.69 &  13.8
 &  228.02 & 13.6 & 230.06 & 13.7 & 228.37 & 13.6 & 230.73 &  13.7 & 229.02  \\ 
 $^{58}$Fe+$^{244}$Pu & 13.8 & 240.88 & 13.7 & 242.62 & 13.7 & 243.34 & 13.8
  & 241.53  & 13.7 & 243.69 & 13.8 & 241.89 & 13.6 & 244.43 &  13.7 & 242.57 \\ 
 $^{64}$Ni+$^{235}$U & 13.8 & 254.33 & 13.7 & 256.17 & 13.7 & 256.97 & 13.8
  &  255.01 & 13.7 & 257.22 & 13.8 & 255.33 & 13.7 & 257.98 &  13.7 & 256.07 \\ 
 $^{66}$Ni+$^{236}$U & 13.9 & 252.59 & 13.8 & 254.33 & 13.8 & 255.06 & 13.9
  &  253.23 & 13.7 & 255.30 & 13.8 & 253.52 & 13.7 & 256.08 & 13.8  & 254.21 \\ 
 $^{50}$Ti+$^{254}$Cf & 13.8  & 212.65  & 13.7 & 214.18  & 13.7  & 214.74 & 13.8
  &  213.22 & 13.7 & 215.15 & 13.7 & 213.57 & 13.6 & 215.74  & 13.7 & 214.18 \\
 $^{54}$Cr+$^{250}$Cm & 13.8  & 226.68 & 13.7 & 228.25 & 13.7 & 228.89   & 13.8
  &  227.26 & 13.7  & 229.28 & 13.8 & 227.61 & 13.6 & 229.90 & 13.7 & 228.22  \\
 $^{60}$Fe+$^{244}$Pu & 13.9  & 239.54  & 13.8 & 241.25  & 13.8  & 241.98 & 13.9
  & 240.17 & 13.7 & 242.27 & 13.8 & 240.51 & 13.7 & 243.05 & 13.8  & 241.19  \\
 $^{72}$Zn+$^{232}$Th & 14.0  & 263.01  & 13.9 & 264.78 & 13.9 & 265.53 &  13.9
 &  263.65 & 13.8 & 265.71 & 13.9 & 263.90 & 13.8 & 266.52  & 13.9 & 264.60  \\
 $^{76}$Ge+$^{228}$Ra & 14.0  & 274.19  & 13.9 & 276.06  & 13.9  & 276.89 & 14.0  & 274.87  & 13.9 & 277.05 & 13.9 & 275.12 & 13.8 & 277.93 & 13.9  & 275.87 \\
\hline
\end{tabular}}
\end{table*}

\subsection{Fixing the relativistic R3Y NN interaction and varying the density:} After comparison of the barrier characteristics and fusion and/or capture cross-section obtained from non-relativistic M3Y and relativistic R3Y and DDR3Y NN interaction potentials, next, we investigate the effects of RMF nuclear density distributions obtained from different force parameter sets on fusion characteristics. In Fig. \ref{fig2}, one notice a small difference at the surface region of interacting nuclei among the nuclear densities given by NL1, NL3, TM1, and DDME2 parameter sets. Since nuclear fusion is a surface phenomenon, the tail region of density distributions plays the most crucial role in the fusion cross-section \cite{raj07}. To study the effect of density distributions on the nuclear potential and consequently on the fusion characteristics, we have fixed the effective NN interaction in the double folding approach and then changed the densities of the fusing nuclei. First, we fixed relativistic R3Y NN potential for the NL1 parameter set and folded it with the density distributions obtained for NL1, NL3, TM1, and DDME2 parameter sets to estimate the nuclear potential from Eq. (\ref{fold}). Then the same procedure is repeated for the R3Y NN potential obtained for the NL3 parameter set. The total interaction potential in terms of barrier height and position, and fusion and/or capture cross-section are then investigated for different density distributions. \\ \\
{\bf Total Interaction potential}: The total interaction potential is calculated for all the nuclear potentials using the same procedure described in the previous subsection. The barrier region of the total interaction potential as a function of radial separation is represented in Fig. \ref{fig6} for the systems (a) $^{48}$Ca+$^{248}$Cm, (b) $^{26}$Mg+$^{248}$Cm and (c) $^{58}$Fe+$^{244}$Pu. The values of $V_B$ and $R_B$ for all the 24 reaction systems are given in Table \ref{tab1}. Here NL3-TM1 signifies that the R3Y NN interaction potential obtained for the NL3 parameter set is folded with the RMF density distributions obtained for the TM1 parameter set. Similarly, NL1-NL3 signifies that the R3Y NN interaction potential obtained for the NL1 parameter set is folded with the RMF density distributions obtained for the NL3 parameter set. Same notations will be used in all the figures and their discussion from here on-wards. The inspection of Fig. \ref{fig6} and Table \ref{tab1} shows that out of eight possible nuclear potentials, the ones with TM1-NL3 and NL1-NL1 give the highest and lowest fusion barriers, respectively. Comparing the characteristics of the barrier for NL1-NL1, NL1-NL3, NL1-TM1, and NL1-DDME2 with the same effective NN interaction potential, we find the densities obtained for NL3, TM1 and DDME2 parameter sets raise the fusion barrier as compared to NL1. The barrier height increases by $\approx 1$ MeV, and the barrier position shifts by $\approx 0.1$ $fm$ towards the lower radial distance as we replace the NL1 densities with those of NL3. Similar behaviour is also observed for the nuclear potentials for NL3-NL3, NL3-NL1, NL3-TM1, and NL3-DDME2. In nutshell, the density distribution obtained for the TM1 parameter set gives the highest fusion barrier, and those for the NL1 parameter set give the lowest fusion barrier. The density distributions given by NL1 parameter sets were observed to be more extended in the surface region than NL3, TM1 and DDME2 densities. It can be inferred from here that a small increase in the densities of fusing nuclei at the surface region lowers the barrier height by approximately 1 MeV. The suppression of barrier height will enhance fusion and/or capture cross-section around the barrier centre of mass energies. To study the effect of density distributions more clearly further, we have calculated the fusion and/or capture cross-section using all six nuclear potentials given in Fig. \ref{fig6} and listed in Table \ref{tab1}. \\ \\
{\bf Fusion and/or capture Cross-Section:} 
To investigate the effects of RMF nuclear density distributions obtained from different parameter sets on the fusion mechanism, next, the fusion and/or capture cross-section for systems (a) $^{48}$Ca+$^{248}$Cm, (b) $^{26}$Mg+$^{248}$Cm and (c) $^{58}$Fe+$^{244}$Pu is calculated within the $\ell$-summed Wong model using nuclear potentials calculated by eight combinations of NN-interaction and RMF density namely NL1-NL1, NL1-NL3, NL1-TM1, NL1-DDME2, NL3-NL1, NL3-NL3, NL3-TM1 and NL3-DDME2 are plotted in Fig. \ref{fig7} as function of center of mass-energy $E_{c.m.}$ (MeV). The available experimental data (black spheres) \cite{itki04,prok05} is also given for the comparison. Comparison of the cross-section calculated using different nuclear potentials shows that the TM1 density distributions decrease the cross-section, whereas the NL1 densities increase the fusion and/or capture cross-section. Also, the difference between the cross-section increases due to different nuclear potential at lower barrier energies. It becomes more prominent as we increase the mass number of the projectile nuclei. For $^{26}$Mg+$^{248}$Cm (Fig. \ref{fig7}(b)) system, the plots of the cross-section for different nuclear potential almost overlap with each other, whereas a larger difference is observed among for $^{58}$Fe+$^{244}$Pu (Fig. \ref{fig7}(c)) system resulting in the formation of SHN Z=120. This shows that the structure effects become more and more crucial as we move towards the heavier mass region of the Periodic Table. Comparison between the experimental and theoretical data shows that for $^{26}$Mg+$^{248}$Cm system, the NL1-NL1 combination is observed to be more suitable than the others. However, for the case of $^{48}$Ca+$^{248}$Cm, a nice fit is observed for NL3-TM1 with the experimental cross-section. All these observations indicate that even a small difference in the density distributions at the surface region significantly impacts the fusion and/or capture cross-section. Also, this effect becomes more prominent as we move towards the superheavy region of the nuclear chart.

\begin{figure*}
\centering
\includegraphics[scale=0.76]{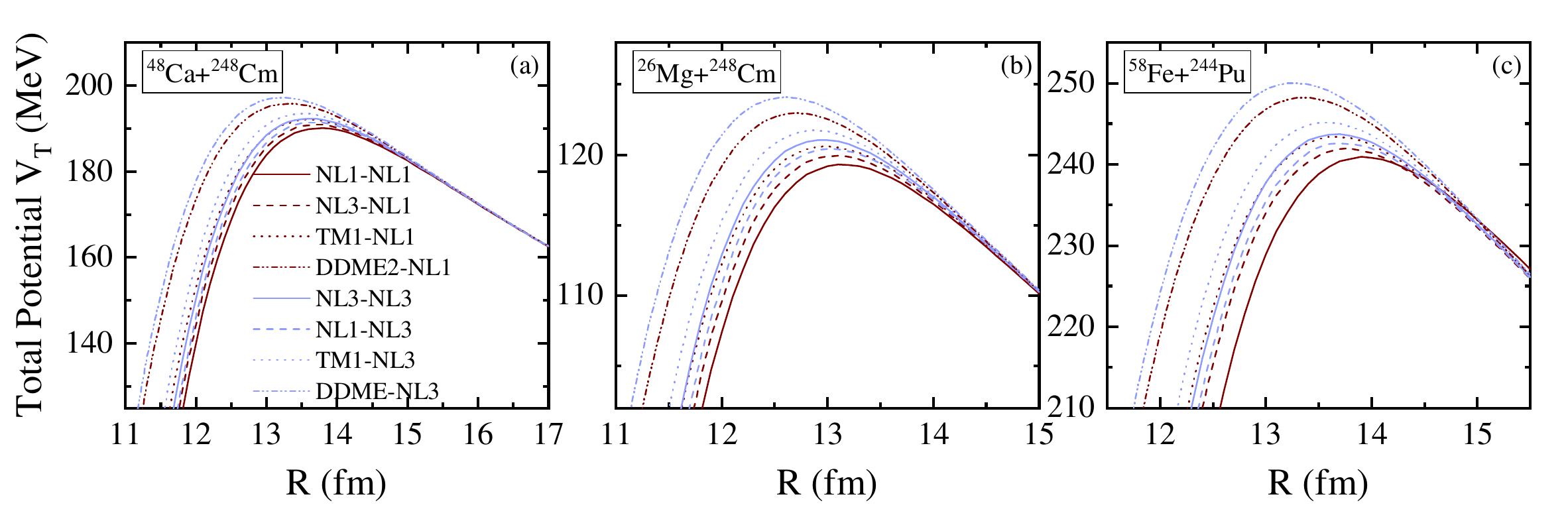}
\caption{(Color online) The variation of total potential $V_T$ (in MeV) as a function of radial distance R (in $fm$) for (a) $^{48}$Ca+$^{248}$Cm, (b) $^{26}$Mg+$^{248}$Cm and (c) $^{58}$Fe+$^{244}$Pu calculated by fixing density for NL1 and NL3 parameter sets. The presentation NL3-NL1 signifies R3Y NN Potential potential from the NL3 parameter set, and density distributions from NL1 parameters are used in the folding procedure to obtain the nuclear potential. The same procedure is followed for other cases as labeled in the figure.} 
\label{fig8}
\end{figure*}
\begin{figure*}
\centering
\includegraphics[scale=0.90]{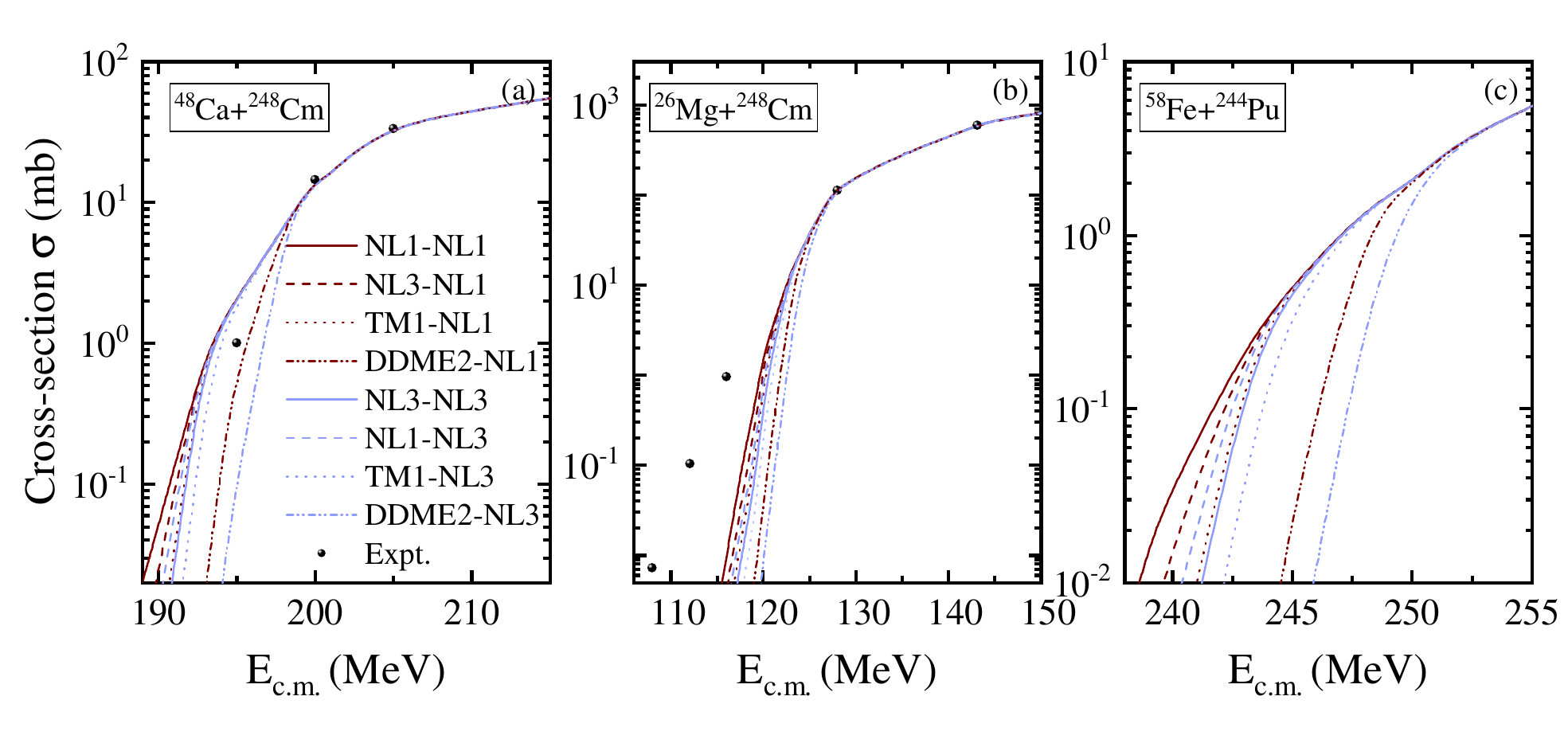}
\caption{(Color online) The cross-section $\sigma$ (mb) calculated with $\ell$-summed Wong model using R3Y NN potential for (a) $^{48}$Ca+$^{248}$Cm, (b) $^{26}$Mg+$^{248}$Cm and (c) $^{58}$Fe+$^{244}$Pu. First, NL1 density is folded with all the four parameter sets (NL1, NL3, TM1 and DDME2), and then the procedure is repeated for NL3 density. The NL3-NL1 signifies R3Y NN Potential potential from the NL3 parameter set and density distributions from NL1 parameters within the folding procedure are used to obtain the nuclear potential. The same procedure is followed for other cases as labeled in the figure.} 
\label{fig9}
\end{figure*}
\subsection{Fixing Density and varying R3Y NN potential}
The double folding optical potential depends upon the nuclear density distributions and the effective nucleon-nucleon interaction. In the previous subsection, we investigated the effect of different nuclear density distributions on the optical nuclear potential and, consequently, on the fusion characteristics. To examine the effects of nucleon-nucleon (NN) interaction on the fusion barrier characteristics, we further fixed the nuclear densities and then changed the effective NN interaction in the double folding approach and studied the fusion characteristics. First, the R3Y NN interaction potential obtained for NL1, NL3 and TM1 parameter sets and DDR3Y NN potential obtained for DDME2 parameter set is integrated over the nuclear densities obtained for the NL1 parameter set. Then the same procedure is repeated with densities obtained for the NL3 parameter set. Again we get eight nuclear potentials denoted as NL1-NL1, NL3-NL1, TM1-NL1, DDME2-NL1, NL1-NL3, NL3-NL3, TM1-NL3, and DDME2-NL3. All the notations have the same meanings as mentioned in the previous subsection. The calculations for the total interaction potential and the fusion and/or capture cross-section are then carried out using these nuclear potentials.

\begin{table}
\caption{\label{tab2} The barrier position $R_B$ (in fm) and barrier height $V_B$ (in MeV) for all the considered 24 reactions by fixing the nuclear density distributions and varying the NN interaction. The TM1-NL1 signifies that the R3Y NN potential obtained for TM1 parameter set are folded with RMF densities obtained for NL1 parameter set.}
\adjustbox{width=8.5cm}{
\begin{tabular}{lllllllll}
\hline
 \multicolumn{1}{c}{\multirow{2}{*}{System}}                              & \multicolumn{2}{c}{TM1-NL1}                               & \multicolumn{2}{c}{DDME2-NL1}                               & \multicolumn{2}{c}{TM1-NL3}                               & \multicolumn{2}{c}{DDME2-NL3}                               \\ \cline{2-9}
 \multicolumn{1}{c}{}                          & \multicolumn{1}{c}{$R_B$} & \multicolumn{1}{c}{$V_B$} & \multicolumn{1}{c}{$R_B$} & \multicolumn{1}{c}{$V_B$} & \multicolumn{1}{c}{$R_B$} & \multicolumn{1}{c}{$V_B$} &  \multicolumn{1}{c}{$R_B$} & \multicolumn{1}{c}{$V_B$} \\ \cline{1-9}
 $^{31}$Al+$^{197}$Au & 12.7 & 109.85 & 12.4  & 112.07  & 12.6 & 110.79 & 12.3 & 113.00  \\ 
$^{39}$K+$^{181}$Ta   & 12.6 & 149.24 & 12.3  & 152.39  & 12.5 & 150.37  & 12.2 & 153.54  \\ 
$^{46}$K+$^{181}$Ta   & 12.9 & 146.19 &  12.6 & 149.21  & 12.8 & 147.34  &  12.5 & 150.36 \\ 
 $^{64}$Ni+$^{238}$U  & 13.7 & 255.73 &  13.4 & 260.91  & 13.6 & 257.55  & 13.3 & 262.76 \\ 
 $^{48}$Ca+$^{238}$U & 13.5 & 185.78 & 13.2  & 189.40 & 13.4 & 187.18  & 13.1 & 190.84 \\ 
  $^{48}$Ca+$^{154}$Sm & 12.5 & 134.92 &  12.2  & 137.55 & 12.4 & 136.13  & 12.1 & 138.77 \\ 
 $^{48}$Ca+$^{248}$Cm    & 13.6 & 192.07 & 13.3 &  195.78  & 13.5 & 193.47 & 13.2  & 197.20   \\ 
 $^{26}$Mg+$^{248}$Cm  & 13.0 & 120.62 & 12.7  & 122.97 & 12.9 & 121.75  & 12.6  & 124.10 \\ 
  $^{40}$Ca+$^{257}$Fm & 13.4 & 202.77 &  13.1  & 206.64 & 13.3 & 204.15  & 13.0 & 208.04 \\ 
 $^{46}$Ti+$^{248}$Cf & 13.4 & 218.93 &  13.1 & 223.15  & 13.3 & 220.54  & 13.0 & 224.78 \\ 
 $^{46}$Ti+$^{249}$Cf & 13.4 & 218.55 & 13.1  & 222.77  & 13.3 & 220.15 & 13.0 &  224.38  \\ 
$^{48}$Ca+$^{254}$Fm & 13.6 & 199.62 &  13.3 & 203.42 & 13.5 & 201.07  & 13.3 & 204.89 \\ 
 $^{50}$Cr+$^{242}$Cm  & 13.4 & 234.56 &  13.1 & 239.13  & 13.3 & 236.37  & 13.0 & 240.96  \\ 
 $^{50}$Ti+$^{249}$Cf & 13.5 & 216.62 & 13.3  & 220.79 & 13.5 & 218.24  & 13.2 & 222.43 \\ 
 $^{50}$Ti+$^{252}$Cf & 13.6 & 215.56 &  13.3 &  219.71 & 13.5 & 217.15  & 13.2 & 221.31  \\ 
 $^{54}$Cr+$^{248}$Cm & 13.6 & 229.76 &  13.3 &  234.28 & 13.5 & 231.43  & 13.3 & 235.95  \\ 
 $^{58}$Fe+$^{244}$Pu  & 13.7 & 243.37 & 13.4  & 248.19  & 13.6 & 245.15   & 13.3 & 249.99  \\ 
 $^{64}$Ni+$^{235}$U & 13.6 & 256.91 & 13.3 &  262.12  & 13.6 & 258.79 & 13.2 & 264.01   \\ 
 $^{66}$Ni+$^{236}$U & 13.7 & 255.14 & 13.4  & 260.41 & 13.6 & 256.90   & 13.3 & 262.22 \\
  $^{50}$Ti+$^{254}$Cf   & 13.7   & 214.85 & 13.4  &  218.99 & 13.6 & 216.47  & 13.3 & 220.57 \\
 $^{54}$Cr+$^{250}$Cm  & 13.7  & 229.01 & 13.4  & 233.51  & 13.6 & 230.66   & 13.3 & 235.17  \\
$^{60}$Fe+$^{244}$Pu   & 13.7  & 241.98 &  13.4 & 246.81  & 13.7 & 243.72  & 13.3  &  248.55 \\
 $^{72}$Zn+$^{232}$Th   & 13.8  & 265.64 & 13.5  & 271.33  & 13.7 & 267.42   & 13.4  & 273.14 \\
$^{76}$Ge+$^{228}$Ra  & 13.8  & 276.92 & 13.5 & 282.85  & 13.7 & 278.81  & 13.4 & 284.75 \\
\hline
\end{tabular}}
\end{table}

\noindent
{\bf Total Interaction Potential:} Fig. \ref{fig8} displays the barrier region of the total interaction potential at $\ell =0\hbar$ as a function of radial separation. As observed in the previous sub-section, it is found here again that the NL1-NL1 combination gives the lowest value of the fusion barrier. The highest barrier is observed for the DDME2-NL3 combination, where the R3Y NN potential is density-dependent. The inclusion of in-medium effects in microscopic R3Y NN potential in terms of density-dependent nucleon-meson coupling parameters is observed to significantly raise the potential barrier. Also, the relativistic R3Y NN potential is calculated with the TM1 parameter which accounts for the isoscalar vector $\omega-$meson's self-coupling is observed to increase the potential barrier height compared to the NL3 and NL1 parameter sets. It is worth mentioning that modification in the fusion barrier height caused by changing the effective NN interaction is comparatively more significant than the one observed by changing the density distributions. The values of barrier heights and positions for all the considered 24 reactions systems are listed in Table \ref{tab2}. On replacing the R3Y NN potential obtained for NL1 with that obtained for TM1 for a fixed density distribution, the barrier height increases by $\approx 1.5$ MeV. The barrier height further increases by upto 5 MeV on replacing the R3Y NN potential obtained for TM1 parameter with DDR3Y NN potential obtained for DDME2 parameter set using relaxed density approximation. This difference in the barrier heights given by R3Y and DDR3Y NN potential is slightly more when folded with NL3 density as compared with NL1 density. Also the change in the barrier characteristics w.r.t. RMF parameter sets become more significant as the mass of the compound nucleus increases. For further exploration of the consequences of different relativistic force parameter sets in terms of NN potential, the fusion and/or capture cross-section for all eight combinations of nuclear potential given in Fig. \ref{fig8} is investigated. \\ \\
{\bf Fusion and/or capture Cross-Section:} The fusion and/or capture cross-section for the systems (a) $^{48}$Ca+$^{248}$Cm, (b) $^{26}$Mg+$^{248}$Cm and (c) $^{58}$Fe+$^{244}$Pu is obtained as a function of center-of-mass energy and is represented in Fig. \ref{fig9}. The effect of the interaction potential characteristics are observed directly in the fusion and/or capture cross-section. We obtained the highest cross-section for the combination NL1-NL1, whereas the lowest cross-section was observed for the DDME2-NL3 parameter set. Also, the DDR3Y NN potential obtained for the DDME2 parameter set gives the lower cross-section as compared to the medium-independent R3Y NN potential obtained for non-linear NL1, NL3, and TM1 parameter sets.  This indicates that the inclusion of in-medium effects in the microscopic NN potential decreases the cross-section. The structure effects of the interaction potential are observed to be diminished at energies greater than the fusion barrier. For system $^{48}$Ca+$^{248}$Cm (Fig. \ref{fig9}(a)), the DDR3Y NN potential obtained for the DDME2 parameter set folded with NL1 density give a better fit to the experimental data. However for the system with lighter projectile i.e. $^{26}$Mg+$^{248}$Cm (Fig. \ref{fig9}(b)), the parameter set NL1 gives better results. The effects of varying the effective NN interaction are observed to be more prominent than the nuclear density distributions.

Comparing the barrier characteristics and the fusion and/or capture cross-section obtained for M3Y and R3Y NN potential folded with nuclear density distributions for four different parameters sets, it is concluded that relativistic R3Y NN potentials give a better prediction to the experimental data. Also, as compared to the TM1 and DDME2 parameter set, which were introduced to include the self-coupling of the vector $\omega-$ mesons and density dependence of nucleon-meson couplings, respectively \cite{bodmer91,lala05}, the NL1 and NL3 parameter sets give better overlap with the experimental data for the fusion and/or capture cross-section. Moreover, NL1 is superior to NL3 in addressing the experimental cross-sections. However, in the case of nuclear matter properties, the NL1 parameter set produces a large value of the asymmetry parameter \cite{rein86,lala97}. Moreover NL1 parameter set also fails to fit the neutron skin thickness of the nuclei away from $\beta-$ stability line \cite{afan96,afan96A}. On the other hand, the parameter set NL3 improves the value of the asymmetry parameter without increasing the number of phenomenological parameters. In the present study, the NL3 parameter set gives a better fit to the fusion and/or capture cross-section as compared to the TM1 parameter set, which includes the self-coupling terms of vector meson in RMF Lagrangian \cite{bodmer91}. Taking all these facts into count, it can be concluded that the parameter set NL3 is suitable for describing the bulk and fusion characteristics of finite nuclei (including heavy and superheavy nuclei with higher $N/Z$ ratio) properties of infinite nuclear matter. The TM1 parameter set, which was introduced to incorporate the vector self-coupling to soften the equation of state of nuclear matter \cite{bodmer91}, gives comparatively repulsive nuclear potential in terms of nuclear density distributions and effective NN interaction potential, which consequently underestimate the fusion and/or capture cross-section. The inclusion of density dependence in the R3Y NN potential within the relativistic-Hartree-Bogoliubov approach for the DDME2 parameter set is observed to decrease the cross-section w.r.t. density-independent R3Y NN potentials obtained for NL1, NL3 and TM1 parameter sets. In the present analysis, we have considered only the isospin asymmetric reaction systems i.e. target-projectile combinations forming a neutron-rich compound nucleus. The systematic study for isospin symmetric (N=Z) reaction systems will be carried out in near future.

\section{SUMMARY AND CONCLUSIONS} 
\label{summary}
A systematic study is carried out in order to study the effect of nuclear density distributions and the effective NN interaction on the fusion barrier characteristics. The fusion barrier properties and cross-section of 24 different target-projectile combinations containing the even-even $^{48}$Ca+$^{154}$Sm, $^{48}$Ca+$^{238}$U, $^{48}$Ca+$^{248}$Cm, $^{26}$Mg+$^{248}$Cm; even-odd $^{46}$K+$^{181}$Ta; and odd-odd $^{31}$Al+$^{197}$Au and $^{39}$K+$^{181}$Ta systems as well as systems leading to the synthesis of superheavy isotopes of $Z$ = 120 $^{40}$Ca + $^{257}$Fm, $^{48}$Ca + $^{254}$Fm, $^{46}$Ti + $^{248}$Cf, $^{46}$Ti + $^{249}$Cf, $^{50}$Ti + $^{249}$Cf, $^{50}$Ti + $^{252}$Cf, $^{50}$Cr + $^{242}$Cm, $^{54}$Cr + $^{248}$Cm, $^{58}$Fe + $^{244}$Pu, $^{64}$Ni+$^{238}$U, $^{64}$Ni + $^{235}$U, $^{66}$Ni + $^{236}$U, $^{50}$Ti + $^{254}$Cf, $^{54}$Cr + $^{250}$Cm, $^{60}$Fe + $^{244}$Pu, $^{72}$Zn + $^{232}$Th and $^{76}$Ge + $^{228}$Ra are investigated within the relativistic mean field formalism. The nuclear density distributions for all the interacting nuclei are obtained by employing the RMF formalism for NL1, NL3, TM1 parameter sets and the RHB approach for the DDME2 parameter set. The effective NN interactions are obtained using the well-known M3Y potential, the relativistic R3Y and DDR3Y potentials. The R3Y NN potential is obtained for the considered three relativistic mean-field parameter sets (NL1, NL3 and TM1) and DDR3Y NN potential is obtained within the RHB approach for the DDME2 parameter set.

In the first step, the comparison of M3Y, R3Y and DDR3Y NN potentials is carried out by calculating the fusion barrier characteristics using the nuclear potential within the double folding approach. It is concluded that the relativistic R3Y and DDR3Y NN interaction potentials give a relatively better fit to the experimental data than the M3Y potential. It is also observed that the NL1 and NL3 parameter sets give a better fit to the experimental data than the TM1 and DDME2 parameter sets. However for systems $^{46}$K+$^{181}$Ta, $^{48}$Ca+$^{248}$Cm and $^{64}$Ni+$^{238}$U, TM1 parameter set works better. Secondly, the effective NN interaction is fixed, and then the nuclear density distributions are changed in the folding procedure to study their effect on the fusion characteristics. It is noticed that the nuclear densities obtained for parameter sets NL3, TM1 and DDME2 give comparatively repulsive nuclear potential and consequently decrease the fusion and/or capture cross-section. In the last step, we studied the effects of effective NN interaction on the fusion characteristics by fixing the nuclear densities. We find that the TM1 parameter set gives a repulsive R3Y NN interaction potential and thus decreases the fusion probability. All these observations lead to the conclusion that the inclusion of vector self-coupling term ($\propto \omega^4$) in RMF Lagrangian increases the magnitude of the repulsive core of the NN interaction, which consequently underestimates the cross-section. Moreover, the DDR3Y NN potential calculated in terms of density-dependent nucleon-meson couplings within the relativistic-Hartree-Bogoliubov (RHB) approach for DDME2 parameter set is observed to give the higher barrier height and lower fusion cross-section as compared to the medium-independent R3Y NN potential obtained within RMF formalism for NL1, NL3 and TM1 parameter sets. From the comparison of the theoretical cross-section with the available experimental data, it is concluded that both the densities and R3Y NN potential obtained for NL1 and NL3 parameters give comparatively better overlap than the TM1 and DDME2 parameter sets. However, if we consider the overall description of bulk properties and fusion characteristics of finite nuclei and the properties of infinite nuclear matter, the NL3 becomes the favourable choice. It is worth mentioning that the shape degrees of freedom, i.e., the nuclear deformation for the interacting nuclei, are not considered in the present analysis. Hence, the results may change slightly without affecting the predictions by including nuclear shape degrees of freedom within the relativistic mean-field formalism, which will be carried out shortly.

\section*{Acknowledgements} \noindent
This work has been supported by Board of Research in Nuclear Sciences (BRNS), Department of Atomic Energy (DAE), Govt. of India, Sanction No. 58/14/12/2019-BRNS, Science
Engineering Research Board (SERB), File No. CRG/2021/001229, FOSTECT Project Code: FOSTECT.2019B.04 and FAPESP Project Nos. 2017/05660-0. 


\end{document}